\documentclass[vecphys]{svmult}

\usepackage{makeidx}       
\usepackage{graphicx}       
\usepackage{times}
\usepackage{multicol}        
\usepackage[bottom]{footmisc}

\makeindex             

\begin{document}

\title*{THE COUPLED ELECTRON-ION MONTE CARLO METHOD}

\author{Carlo Pierleoni\inst{1} \and David M. Ceperley\inst{2}}

\institute{Department of Physics, University of L'Aquila,
Polo di Coppito, Via Vetoio, L'Aquila, 67010 Italy, \texttt{carlo.pierleoni@aquila.infn.it}
\and Department of Physics and NCSA, University of Illinois at
Urbana-Champaign, Urbana, IL 61801, USA, \texttt{david.ceperley@uiuc.edu}}

\maketitle

%

\section{Introduction}     
Twenty years ago Car and Parrinello introduced an efficient method
to perform Molecular Dynamics simulation for classical nuclei with
forces computed on the ``fly'' by a Density Functional Theory
(DFT) based electronic calculation\cite{cpmd85}. Because the
method allowed study of the statistical mechanics of classical
nuclei with many-body electronic interactions, it opened the way
for the use of simulation methods for realistic systems with an
accuracy well beyond the limits of available effective force
fields. In the last twenty years, the number of applications of
the Car-Parrinello ab-initio molecular dynamics has ranged from
simple covalent bonded solids, to high pressure physics, material
science and biological systems. There have also been extensions of
the original algorithm to simulate systems at constant temperature
and constant pressure\cite{bernasconi95}, finite temperature
effects for the electrons \cite{alavikohanoffpf}, and quantum
nuclei\cite{marx-par}.

DFT is, in principle, an exact theory but the energy functional
are treated approximately at the level of a self consistent mean
field theory for practical purposes. Despite recent progress, DFT
suffers from well-known limitations, for example, excited state
properties such as optical gaps and spectra are generally
unreliable. DFT shows serious deficiencies in describing van der
Waals interactions, non-equilibrium geometries such as reaction
barriers, systems with transition metals and/or cluster isomers
with competing bonding patterns\cite{martin04,Foulkes01}. As a
consequence, current ab-initio predictions of metallization
transitions at high pressures, or even the prediction of phase
transitions are often only qualitative. Hydrogen is an extreme
case \cite{maksimivshilov,stadelemartin,johnsonaschroft} but even
in silicon the diamond/$\beta$-tin transition pressure and the
melting temperature are seriously underestimated\cite{alfe04}.

An alternative route to the ground state properties of a system of
many electrons in presence of nuclei is the Quantum Monte Carlo
method (QMC) \cite{hammond,Foulkes01}. QMC methods for bosons are
``exact '' meaning that all systematic errors are under control
and can be reduced as much as desired with a computational cost
growing as a power of the number of particles. However for
fermions the ``sign problem'' makes a direct extension of QMC
unstable and one has to resort to the ``fixed node approximation''
for practical calculations\cite{hammond,Foulkes01}. Over the
years, the level of accuracy of the fixed node approximation for
simple homogeneous systems, such as $^{3}He$ and the electron gas,
has been systematically improved \cite{panoffcarlson,kwon,hcpe03}.
In more complex, inhomogeneous situations such as atoms, molecules
and extended systems of electrons and nuclei, progress have been
somewhat slower. Nonetheless, in most cases, fixed-node QMC
methods have proved to be more accurate than mean field methods
(HF and DFT)\cite{Foulkes01}. Computing ionic forces with QMC to
replace the DFT forces in the ab-initio MD, is more difficult and
a general and efficient algorithm is still missing. Moreover, the
computer time required for a QMC estimate of the electronic energy
is, in general, more than for a corresponding DFT-LDA calculation.
These problems have seriously limited the development of an
ab-initio simulation method based on the QMC solution of the
electronic problem ``on the fly''.

In recent years, we have developed a different strategy based
entirely on the Monte Carlo method, both for solving the
electronic problem, and for sampling the ionic configuration
space\cite{DewingCeperley01,dmc03}. The new method, called the Coupled
Electron-Ion Monte Carlo method (CEIMC) relies on the
Born-Oppenheimer approximation for treating finite temperature
ions coupled with ground state electrons. A Metropolis Monte Carlo
simulation of the ionic degrees of freedom (represented either by
classical point particles or by path integrals) at fixed
temperature is performed based on the electronic energies computed
during independent ground state Quantum Monte Carlo calculations.
CEIMC has been applied, so far, to high pressure metallic hydrogen
where it has found quite different effects of temperature than
CPMD with Local Density Approximation (LDA) forces\cite{pch04}. In
these lecture notes, we present the theoretical basis of CEIMC. We
start by describing, in some detail, the ground state QMC methods
implemented in CEIMC, namely the Variational Monte Carlo and the
Reptation Quantum Monte Carlo methods. We then describe the fixed
node or restricted paths approximation necessary to treat fermions
and the fixed phase method used to perform the average over
boundary conditions needed for metallic system. In the subsequent section we describe
how to implement the Metropolis algorithm when the energy
difference has a statistical noise and discuss efficient
strategies for energy differences within the CEIMC. The
next section is devoted to describe the method to treat quantum
mechanical protons and to integrate efficiently this new
difficulty within CEIMC. Finally, we briefly review some CEIMC
results for high pressure hydrogen and compare with existing CPMD
results. Conclusions and perspectives for future developments are
collected in the last section.

\section{The electronic ground state problem}

Let us consider a system of $N_{p}$ nuclei and $N_{e}$ electrons
in a volume $V$ described by the non relativistic hamiltonian
\begin{equation}
\hat{H}=-\sum_{i=1}^{N}\lambda_{i}\nabla_{i}^{2} + \frac{e^{2}}{2}\sum_{i\neq j}^{N}
\frac{z_{i}z_{j}}{|\hat{\vec{r}}_{i}-\hat{\vec{r}}_{j}|}
\label{eq:hamiltonian}
\end{equation}
where $N=N_{e}+N_{p}$, and $z_{i}$, $m_{i}$, $\hat{\vec{r}}_{i}$
represent the charge, mass and position operator of particle $i$
respectively and $\lambda_{i}=\hbar^{2}/2m_{i}$. Let us denote
with $R=(\vec{r}_{1},\cdots,\vec{r}_{N_{e}})$ and
$S=(\vec{r}_{N_{e}+1},\cdots,\vec{r}_{N})$ the set of coordinates
of all electrons and nuclei respectively.
We restrict the discussion to unpolarized systems, i. e. systems with a
vanishing projection of the total spin along a given direction,
say $S_{z}=0$.  Since the Hamiltonian does not flip spins, we can
label electrons from $1$ to $N_{e}/2$ as up spin ($\uparrow$) and
electrons from $N_{e}/2+1$ to $N_{e}$ as down spin ($\downarrow$).

Within the Born-Oppenheimer approximation, the energy of the
system for a given nuclear state $S$ is the expectation value
of the hamiltonian $\hat{H}$ over the corresponding exact ground
state $|\Phi_{0}(S)>$
\begin{equation}
E_{BO}(S)={<\Phi_{0}(S)|\hat{H}|\Phi_{0}(S)>}
\end{equation}
which is a $3N_{e}$ dimensional
integral over the electronic coordinates in configurational space
\begin{equation}
E_{BO}(S)=\int dR~\Phi_{0}^{*}(R|S) \hat{H}(R,S) \Phi_{0}(R|S) =
\int dR \left|\Phi_{0}(R|S)\right|^{2} E_{L}(R|S)
\label{eq:Ebo}
\end{equation}
with the \textit{local energy} defined as
\begin{equation}
E_{L}(R|S)=\frac{\hat{H}(R,S)\Phi_{0}(R|S)}{\Phi_{0}(R|S)}
\label{eq:Eloc}
\end{equation}
Since $\Phi_{0}(R|S)$ is normalized and
$\left|\Phi_{0}(R|S)\right|^{2}\geq0$ everywhere, the $3N_{e}$
dimensional integral in eq. (\ref{eq:Ebo}) can be performed by
standard Metropolis Monte Carlo by generating a Markov process
which sample asymptotically $\left|\Phi_{0}(R|S)\right|^{2}$.
Expectation values of any observable can be computed along the
same Markov chain. In this respect, computing the properties of a
many-body quantum system is similar to performing a MC calculation
for a classical system. The square modulus of the ground state
wave function plays the role of the classical Boltzmann
distribution. An important quantity in what follows is the measure
of the energy fluctuations for a given wave function. This can be
defined by the variance of the local energy
\begin{eqnarray}
\sigma^{2}(S)&=&\int dR \left|\Phi_{0}(R|S)\right|^{2}
\left( \frac{\hat{H}(R,S)\Phi_{0}(R|S)}{\Phi_{0}(R|S)}-E_{BO}(S) \right)^{2}\nonumber \\
&=&\int dR \left(\hat{H}\Phi_{0}(R|S) \right)^{2} - E_{BO}^{2}(S)
\end{eqnarray}
Note that for any exact eigenfunction of the hamiltonian, the
local energy $E_{L}(R|S)$ does not depend on the electronic
configuration $R$ and is equal to the corresponding eigenvalue.
This implies that the variance vanishes. This is known as the zero
variance principle in Quantum Monte Carlo.

\subsection{Variational QMC}
The problem is to get a computable expression for the ground state
wave function without solving the Schr\"odinger equation for the
many body hamiltonian of eq. (\ref{eq:hamiltonian}), obviously an
impossible task for any non trivial system. As usual in many body
problems, we can resort to the variational principle which states
that the energy of any proper trial state $|\Psi_{T}(S)>$ will be
greater or equal to the ground state energy
\begin{equation}
E_{BO}(S) \leq E_{V}(S) = \frac{\int dR~\Psi_{T}^{*}(R|S) \hat{H}(R,S) \Psi_{T}(R|S)}{\int dR~\Psi_{T}^{*}(R|S) \Psi_{T}(R|S)}
\label{eq:Ev}
\end{equation}
A proper trial wave function must satisfy the following
requirements
\begin{itemize}
\item it has to have the right symmetry under particle
permutation: $\Psi_{T}(\hat{P}R|S)=(-1)^{P}\Psi_{T}(R|S)$, where
$\hat{P}$ is the permutation operator for electrons of same spin.
\item the quantity $\hat{H}\Psi_{T}$ needs to be well defined
everywhere which implies that both $\Psi_{T}$ and $\nabla\Psi_{T}$
must be continuous whenever the potential is finite, including at
the periodic boundaries. \item the integrals $\int dR
|\Psi_{T}|^{2}$ and $\int dR \Psi_{T}^{*}\hat{H}\Psi_{T}$ must
exist. Furthermore for a Monte Carlo evaluation of the variance
$\sigma^{2}$ the integral $\int dR (\hat{H}\Psi_{T})^{2}$is also
required to exist.
\end{itemize}
For a given trial function, it is essential to show analytically
that these properties hold everywhere, in particular at the edge
of the periodic box and when two particles approach each other
(where generally the potential diverges). Otherwise, either the
upper bound property is not guaranteed or the Monte Carlo error
estimates are not valid.\\
The strategy in Variational Monte Carlo (VMC) is therefore to pick
a proper form for a trial wave function based on physical insight
for the particular system under study. In general, a number of
parameters $(\alpha_{1},\cdots,\alpha_{k})$ will appear in the
wave function to be treated as variational parameters. For any
given set of $\{\alpha\}$ the Metropolis algorithm is used to
sample the distribution
\begin{equation}
\Pi(R|S,\{\alpha\})=\frac{\Psi_{T}(R|S,\{\alpha\})}{\int dR \Psi_{T}(R|S,\{\alpha\})}
\end{equation}
and the electronic properties are then computed as averages over the generated Markov chain
\begin{equation}
E_{V}^{T}(S,\{\alpha\})=<E_{L}(R|S,\{\alpha\})>
\end{equation}
where the superscript $T$ as been explicitly
written to remember that the variational energy depends in general
on the chosen analytical form and, for any given form, on the numerical values
of the variational parameters $\{\alpha\}$.\\
Because of the zero variance principle stated above, the
fluctuations in the local energy are entirely due to inaccuracies
of the trial function for the particular configurations generated
during the MC run. As the trial wave function approaches the exact
eigenfunction (everywhere in configuration space!) the
fluctuations decrease and the variational estimate of the energy
converges more rapidly with the number of MC steps. At the same
time, the estimate converges to the exact energy. This is at
variance with classical Monte Carlo where fluctuations are induced
by temperature.

The variational method is very powerful, and intuitively pleasing.  One posits
a form of the trial function and then obtains an upper bound for the energy. In contrast
to other theoretical methods, no further approximations are
made. The only restriction on the trial function
is to be computable in a reasonable amount of time.

One of the problems with VMC is that it favors simple states over
more complicated states. As an example, consider the liquid-solid
transition in helium at zero temperature.  The solid wave function
is simpler than the liquid wave function because in the solid the
particles are localized so that the phase space that the atoms
explore is much reduced.  This biases the difference between the
liquid and solid variational energies for the same type of trial
function, ({ \it e.g.} a pair product form, see below) since the
solid energy will be closer to the exact result than the liquid.
Hence, the transition density will be systematically lower than
the experimental value.  Another illustration is the calculation
of the polarization energy of liquid $~^3$He. The wave function
for fully polarized helium is simpler than for unpolarized helium
because antisymmetry requirements are higher in the polarized
phase so that the spin susceptibility computed at the pair product
level has the wrong sign!

The optimization of trial functions for many-body systems is time
consuming, particularly for complex trial functions.
The dimension of the parameter space increases rapidly with the
complexity of the system and the optimization can become very
cumbersome since it is, in general, a nonlinear optimization
problem. Here we are not speaking of the computer time, but of the
human time to decide which terms to add, to program them and their
derivatives in the VMC code. This allows an element of human bias
into VMC; the VMC optimization is more likely to be stopped when
the expected result is obtained. The basis set problem is still
plaguing quantum chemistry even at the SCF level where one only
has 1-body orbitals. VMC shares this difficulty with basis sets as
the problems get more complex.

Finally, the variational energy is insensitive to long range order.  The energy
is dominated by the local order (nearest neighbor correlation functions).
If one is trying to compare the variational energy of a trial function
with and without long range order, it is extremely important that both
functions have the same short-range flexibility and both trial functions
are equally optimized locally.
Only if this is done, can one have any hope of saying anything about the
long range order. The error in the variational energy is second order in
the trial function, while any other property will be first order.  Thus
variational energies can be quite accurate while correlation functions are
not very accurate.

As a consequence, the results typically reflect what was put into
the trial function. Consider calculating the momentum
distribution. Suppose the trial function has a Fermi surface. Then
the momentum distribution
will exhibits a discontinuity at $k_{f}$ signaling the presence of a Fermi surface.
This does not imply that the true wave function has a sharp Fermi surface.

\subsection{Reptation Quantum Monte Carlo}
It is possible to go beyond VMC by a number of related methods
known as Projection Monte Carlo Methods. The general idea is to
chose a trial function which has a non negligible overlap with the
ground state wave function (in general, it is enough to require
the right symmetry) and to apply a suitable projection operator
which zeros out all the components of the trial wave function from
the excited states in the Hilbert space of the system.
We will limit the description to the method implemented in CEIMC,
namely the Reptation Quantum Monte Carlo \cite{BaroniMoroni} or Variational Path Integral \cite{rmp95,SarsaSchmidtMagro00}.
For the discussion of other projection QMC methods such as Diffusion Monte Carlo (DMC)
and Green Function Monte Carlo (GFMC), we refer to the specialized
literature\cite{Foulkes01,hammond}.\\
Let us define $\{\Phi_{i},E_{i}\}$ as the complete set of
eigenfunction and eigenvalues of the hamiltonian $\hat{H}$ in
eq.(\ref{eq:hamiltonian}). Any trial state can be decomposed in
the eigenstate basis:
\begin{equation}
|\Psi_{T}>= \sum_{i} c_{i} |\Phi_{i}>
\end{equation}
where $c_{i}$ is the overlap of the trial state with the $i^{th}$ eigenstate.
Let us consider the application of the operator $e^{-t\hat{H}}$ onto this state
\begin{equation}
|\Psi(t)>=e^{-t\hat{H}}|\Psi(0)>=\sum_{i}c_{i}e^{-tE_{i}} |\Phi_{i}>
\end{equation}
with the initial state $|\Psi(0)>=|\Psi_{T}>$. Here $t$ is a
control parameter with dimension of inverse energy and we will
call it ``time'' since it plays the role of imaginary time in the
Bloch equation (see below). All excited states will be zeroed
exponentially fast with increasing $t$, the rate of the
convergence to the ground state depending on the energy gap
between the ground state and the first excited state
non-orthogonal to the trial function. The total energy as function
of time is defined as
\begin{equation}
E(t)=\frac{<\Psi(t/2)|\hat{H}|\Psi(t/2)>}{<\Psi(t/2)|\Psi(t/2)>}=
\frac{<\Psi_{T}|e^{-\frac{t}{2}\hat{H}}\hat{H}e^{-\frac{t}{2}\hat{H}}|\Psi_{T}>}
{<\Psi_{T}|e^{-t\hat{H}}|\Psi_{T}>}
\label{eq:E(t)}
\end{equation}
Similar to a thermal partition function, let us define the generating
function of the moments of $\hat{H}$ as
\begin{equation}
Z(t)=<\Psi_{T}|e^{-t\hat{H}}|\Psi_{T}>.
\end{equation}
The total energy at time $t$ is simply the derivative of the logarithm of $Z(t)$
\begin{equation}
E(t)=-\frac{\partial}{\partial t}ln Z(t)
\end{equation}
and the variance of the energy is the second derivative
\begin{equation}
\sigma^{2}_{E}(t)= <(\hat{H}-E(t))^{2}>= -\frac{\partial}{\partial t}E(t) \geq 0
\end{equation}
which is non-negative by definition. This implies that the
energy decreases monotonically with time. The ground state is
reached at large time (much larger than the inverse gap) and
\begin{eqnarray}
\lim_{t\rightarrow\infty} E(t)&=&E_{0}\\
\lim_{t\rightarrow\infty} \sigma^{2}(t)&=&0
\label{eq:ZVRQMC}
\end{eqnarray}
The last relation is the generalization of the zero variance principle in Projection Monte Carlo.\\
For observables $\hat{A}$ which do not commute with $\hat{H}$, for instance correlation functions, the average at ``time'' $t$,
defined as in eq. (\ref{eq:E(t)}), takes the following form in configurational space
\footnote{the expression gets slightly easier for observables diagonal in configurational space: $<R|A|R'>=A(R)\delta(R-R')$.}
\begin{eqnarray}
A(t)=<\hat{A}>_{t}=\frac{1}{Z(t)}\int &&dR_{1}dR_{2}dR_{3}dR_{4} <\Psi_{T}|R_{1}> \rho(R_{1},R_{2}|\frac{t}{2})\nonumber \\
& &<R_{2}|\hat{A}|R_{3}>\rho(R_{3},R_{4}|\frac{t}{2})<R_{4}|\Psi_{T}>
\label{eq:A(t)}
\end{eqnarray}
where $R_{i}$ represent the set of all electronic coordinates and $\rho(R,R',t)$
is the thermal density matrix of the system at inverse temperature $t$
\begin{equation}
\rho(R,R',t)=<R|e^{-t\hat{H}}|R'>
\end{equation}
Similarly, the expression of $Z(t)$ in configurational space is
\begin{equation}
Z(t)=\int dR_{1}dR_{2}<\Psi_{T}|R_{1}>\rho(R_{1},R_{2},t)<R_{2}|\Psi_{T}>
\label{eq:GeneratingFunction}
\end{equation}
Thus, in order to compute any average over the ground state we
need to know the thermal density matrix at large enough ``time''.
Obviously, its analytic form for any non-trivial many-body system
is unknown. However, at short time (or high temperature) the
system approaches its classical limit and we can obtain
approximations. Let us first decompose the time interval $t$ in
$M$ smaller time intervals, $\tau=t/M$
\begin{equation}
\rho(R,R',t)=<R|e^{-(\tau\hat{H})^{M}}|R'>=
\int dR_{1}\cdots dR_{M-1}\prod_{k=1}^{M-1}<R_{k-1}|e^{-\tau\hat{H}}|R_{k}>
\end{equation}
with the boundary conditions: $R_{0}=R$ and $R_{M}=R'$ on the
paths. For $M$ large enough, we can apply the Trotter
factorization to get an explicit form for the short time
propagator. The simplest factorization, known as the ``primitive''
approximation, consists of ignoring the commutator of the kinetic
and potential operators
\begin{equation}
\rho(R_{k-1},R_{k},\tau)=<R_{k-1}|e^{-\tau\hat{H}}|R_{k}>
\simeq <R_{k-1}|e^{-\tau\hat{K}}|R_{k}> e^{-\frac{\tau}{2}\left[V(R_{k})+V(R_{k-1})\right]}
\label{eq:primitive}
\end{equation}
A more accurate, but also more complex form, will be discussed
later in the section on Path Integral Monte Carlo. Note that we
have symmetrized the primitive form in order to reduce the
systematic error of the factorization \cite{rmp95}. The explicit
form of the kinetic propagator is the Green's function of the
Bloch equation of a system of free fermions \cite{FeynmanStatMech,rmp95}, i.e. a diffusion
equation in configurational space
\begin{equation}
<R_{k-1}|e^{-\tau\hat{K}}|R_{k}> = \left(\frac{1}{4\pi\lambda\tau}\right)^{\frac{3N}{2}}
e^{-\frac{\left|R_{k}-R_{k-1}\right|^{2}}{4\lambda\tau}}
\end{equation}
and therefore we get
\begin{equation}
\rho(R,R',t)=\int\prod_{k=1}^{M-1}dR_{k}
\left[\prod_{k=1}^{M} \frac{e^{-\frac{\left|R_{k}-R_{k-1}\right|^{2}}{4\lambda\tau}}}
{\left(4\pi\lambda\tau\right)^{3N/2}}\right] e^{-\tau\left[\frac{V(R_{0})}{2}+\sum_{k=1}^{M-1} V(R_{k})+\frac{V(R_{M})}{2}\right]}
\end{equation}
In the continuous limit ($M\rightarrow\infty$, $\tau\rightarrow0$, $t=M\tau=$const.) it becomes the Feynman-Kac formula \cite{rmp95}
\begin{equation}
\rho(R,R',t)=\left<\exp\left(-\int_{0}^{t}d\tau V(R(\tau))\right)\right>_{RW}
\label{eq:FeynmanKac}
\end{equation}
where $<\cdots>_{RW}$ indicate a path average over gaussian random walks $R(\tau)$
starting at $R(0)=R$ and ending at $R(t)=R'$ in a time t. \\
We have, in principle, developed a scheme for Monte Carlo calculations of
ground state averages of a general quantum system.

However, this scheme has a serious problem of efficiency which
prevents its use for any non-trivial system. At the origin of the
problem are the wild variations of the potential $V(R)$ in
configuration space. There are cases like electron-proton systems
where the potential is not bounded and therefore the primitive
approximation of the propagator is not stable for any finite time
step $\tau$. However, even with well behaved effective potentials
(like in Helium for instance) the large fluctuations of the
potential energy would require a very small time step in order to
observe convergence of the averages to their exact value.
Moreover, the efficiency will degrade rapidly with more particles.
The problem was recognized in the early days of QMC and the remedy
introduced by Kalos in 1974. In the community of Ground State QMC it goes under
the name of ``importance sampling'' (IS). A different strategy is applied in the PIMC community.
We now describe importance sampling, not in the original form as introduced by Kalos in
Green Function Monte Carlo, but following a recent development by Baroni and Moroni
in the framework of the Reptation QMC \cite{BaroniMoroni}.\\
In VMC, a good trial function should have a local energy almost
constant in configuration space. Let us assume to know such
function $\Psi_{T}$. Let us then rewrite the hamiltonian $\hat{H}$
in terms of a new fictitious hamiltonian $\hat{\cal{H}}$
\begin{equation}
\hat{H}=\hat{\cal{H}}+E_{L}(R)
\end{equation}
where
\begin{eqnarray}
\hat{\cal{H}}&=& \lambda \left[-\nabla^{2}+\frac{\nabla^{2}\Psi_{T}}{\Psi_{T}}\right] \\
E_{L}(R)&=&V(R)-\lambda\frac{\nabla^{2}\Psi_{T}}{\Psi_{T}}
\end{eqnarray}
We can now factorize the short time propagator in a different way (confr. eq. (\ref{eq:primitive}) )
\begin{equation}
\rho(R_{k-1},R_{k},\tau)
\simeq <R_{k-1}|e^{-\tau\hat{\cal{H}}}|R_{k}> e^{-\frac{\tau}{2}\left[E_{L}(R_{k})+E_{L}(R_{k-1})\right]}
\end{equation}
In this new form, the widely oscillating potential energy is
replaced by the local energy which is much smoother for an
accurate $\Psi_{T}$. We need to find the short time
propagator of the importance sampling hamiltonian $\hat{\cal{H}}$
which is nothing but the solution of the corresponding Bloch
equation  \cite{FeynmanStatMech,rmp95}
\begin{eqnarray}
-{\partial_{t}} \rho_{_{IS}}(R,R',t)&=&\hat{\cal{H}}\rho_{_{IS}}(R,R',t)\\
\rho_{_{IS}}(R,R',0)&=&\delta(R-R')
\label{eq:ISBlochEq}
\end{eqnarray}
It is not difficult to show by direct substitution, that the short
time solution of this equation is
\begin{equation}
\rho_{_{IS}}(R_{k-1},R_{k},\tau)= \frac{\Psi_{T}(R_{k-1})}{\Psi_{T}(R_{k})} \left(\frac{1}{4\pi\lambda\tau}\right)^{\frac{3N}{2}}
\exp\left\{-\frac{(R_{k}-R_{k-1}-2\lambda\tau F_{k-1})^{2}}{4\lambda\tau}\right\}
\label{eq:rhoISns}
\end{equation}
if we make the short time approximation
$\left[1+\tau\left(\nabla F+\frac{1}{F}\nabla^{2}F\right)\simeq 1\right]$.
In these expressions, the drift force is defined as $F_{k}=F(R_{k})=2\nabla_{R_{k}}\ln\Psi_{T}(R_{k})$. \\
This form of the short time propagator does not satisfy an
important property of density matrices, namely the symmetry under
exchange of the two legs, $R$ and $R'$. 
We can remedy by taking the symmetrized density matrix as short
time propagator
\begin{equation}
\rho_{_{IS}}^{s}(R_{k-1},R_{k},\tau)=\left[\rho_{_{IS}}(R_{k-1},R_{k},\tau) \rho_{_{IS}}(R_{k},R_{k-1},\tau)\right]^{\frac{1}{2}}
=\frac{e^{-L_{s}(R_{k-1},R_{k},\tau)}}{\left(4\pi\lambda\tau\right)^{\frac{3N}{2}}}
\label{eq:rhoISs}
\end{equation}
where the expression for the symmetrized link action is
\begin{equation}
L_{s}(R_{k-1},R_{k},\tau)=\frac{(R_{k}-R_{k-1})^{2}}{4\lambda\tau}+\frac{\lambda\tau}{2}(F_{k}^{2}+F_{k-1}^{2})+\frac{(R_{k}-R_{k-1})\cdot(F_{k}-F_{k-1})}{2}
\label{eq:linkaction}
\end{equation}
Using eqs. (\ref{eq:rhoISs}), (\ref{eq:linkaction}) we obtain the propagator at any time $t$ as
\begin{equation}
\rho(R,R',t)=\int \prod_{k=1}^{M-1}dR_{k}
\left[\prod_{k=1}^{M} \frac{e^{-L_{s}(R_{k-1},R_{k},\tau)}}
{\left(4\pi\lambda\tau\right)^{3N/2}}\right]
e^{-\tau\left[\frac{E_{L}(R_{0})}{2}+\sum_{k=1}^{M-1} E_{L}(R_{k})+\frac{E_{L}(R_{M})}{2}\right]}
\label{eq:DiscreteDM}
\end{equation}
In the continuum limit it is the generalized Feynman-Kac
formula
\begin{equation}
\rho(R,R',t)=\left<\exp\left(-\int_{0}^{t}d\tau E_{L}(R(\tau))\right)\right>_{DRW}
\label{eq:GeneralizeFeynmann-Kac}
\end{equation}
where $<\cdots>_{DRW}$ indicate a path average over drifted random walks starting at $R(0)=R$
and ending at $R(t)=R'$ in a time t. With this form of the density matrix the generating
function takes the form
\begin{equation}
Z(t)=\int dR dR' \Psi_{T}(R) \left<e^{-\int_{0}^{t}d\tau E_{L}(Q(\tau))}\right>_{DRW}
\Psi_{T}(R')
\label{eq:ZetaRQMC}
\end{equation}
and the average of a generic observable $\hat{A}$ becomes
\begin{eqnarray}
A(t)=\frac{1}{Z(t)}\int \prod_{k=1}^{4}dR_{k}  \Psi_{T}(R_{1}) &&\left<e^{-\int_{0}^{\frac{t}{2}}d\tau E_{L}(Q(\tau))}\right>_{DRW}A(R_{2},R_{3})\nonumber \\
&&\left<e^{-\int_{0}^{\frac{t}{2}}d\tau E_{L}(Q(\tau))}\right>_{DRW} \Psi_{T}(R_{4})
\end{eqnarray}
with obvious boundary conditions on the path averages.\\
A special word on the calculation of the energy is in order. In
the last equality in eq. (\ref{eq:E(t)}) the hamiltonian operator
in the numerator can be pushed either to the left or to the right
in such a way to operate directly on the trial state.  Remembering
the definition of the local energy, eq. (\ref{eq:Eloc}), we can
write
\begin{eqnarray}
E(t)&=&\frac{1}{Z(t)}\int dR dR' E_{L}(R)\Psi_{T}(R)\rho(R,R',t)\Psi(R')\nonumber \\
&=&\frac{1}{Z(t)}\int dR dR' \Psi_{T}(R)\rho(R,R',t)\Psi(R')E_{L}(R') \nonumber \\
&=&\frac{1}{2}<E_{L}(R)+E_{L}(R')>
\end{eqnarray}
We use the last equality in order to improve the efficiency of the estimator.
When computing the variance we push one $\hat{H}$ operator to the left and the other to the right to obtain
\begin{equation}
\sigma^{2}(t)=\int dR dR' E_{L}(R)\Psi_{T}(R)\rho(R,R',t)\Psi(R')E_{L}(R') - E^{2}(t)
\end{equation}
Then reaching the ground state with vanishing variance means taking paths long enough ($t$ large enough)
for the correlation between the two ends to vanish.
On the other hand the VMC method is obtained for $t=0$ as $\rho(R,R',0)=\delta(R-R')$.

\subsection{Fermions}
Up to now we have tacitly ignored the particle statistics and
derived the formalism as if the particles were distinguishable.
As far as the Hamiltonian
does not depend explicitly on spin, the formalism remains 
valid for fermions if we consider states completely antisymmetric under
particle permutation $|\hat{P}R_{i}>=(-)^{P}|R_{i}>$. The
importance sampling hamiltonian $\hat{\cal{H}}$, defined in
the same way, is symmetric under particle exchange even for
antisymmetric trial states. The only place where we need care is
in the initial condition of the Bloch equation, eq.
(\ref{eq:ISBlochEq}), which must be replaced by a completely
antisymmetric delta function
\begin{equation}
\rho_{_{IS}}(R,R',0)=\mathcal{A} \delta(R-R')=\frac{1}{N!}\sum_{P}(-1)^{P}\delta(R-\hat{P}R')
\end{equation}
Since $[\hat{H},\hat{P}]=0$, the imaginary time evolution preserves the
symmetry and the fermion thermal density matrix takes the form
\begin{equation}
\rho_{_{F}}(R,R',t)=\frac{1}{N!}\sum_{P}(-1)^{P}\rho_{_{D}}(R,\hat{P}R',t)
\label{eq:FermionDM}
\end{equation}
where $\rho_{_{D}}$ is the density matrix of a system of
distinguishable particles derived above (see eq.
(\ref{eq:GeneralizeFeynmann-Kac})). The fermion density matrix
between configurations $R$ and $R'$ at ``time'' $t$ is the sum
over permutations of the density matrix of distinguishable
particles between the initial configurations $R$ and the
permutation of the final configuration $\hat{P}R'$, multiplied by
the sign of the permutation. Each of those density matrices arises
from the sum over all paths with given boundary conditions in time
as expressed by the generalized Feynman-Kac formula. We can
therefore think of a path in the configurational space of
distinguishable particles as an object carrying not only a weight
(given by the exponential of minus the integral of the local
energy along the path) but also a sign fixed by its boundary
conditions in time. The fermion density matrix is the algebraic
sum over all those paths. While $\rho_{_{D}}(R,R',t)\ge 0$ for any
$R'$ and $t$ at given $R$, this property obviously does not apply
to $\rho_{_{F}}(R,R',t)$. This is at the origin of the ``fermion
sign problem'' \cite{como95}. Briefly, the sign problem arises from the fact
that the optimal probability to sample the electronic paths is
the absolute value of the fermion density matrix which,
however, is a bosonic density matrix (symmetric under particle
permutation). With this sampling, the sign of the sampled paths
will be left in the estimator for the averages. The normalization
of any average will be given by the number of sampled positive
paths minus the number of sampled negative paths. Since the
sampling is bosonic, i.e. symmetric, these two numbers will
eventually be equal and the noise on any average will blow up for
a long enough sampling. The fundamental reason behind this
pathology is that the
Hilbert space of any time independent Hamiltonian (with local
interactions) can be divided in the set of symmetric states,
antisymmetric states and states of mixed symmetry. These sets are
disjoint, e. g. any symmetric state is orthogonal to any
antisymmetric one; in principle we cannot extract information for
a fermionic state from a bosonic sampling \cite{HuangStatMech}.

A general solution of the fermion sign problem is still
unavailable, although interesting algorithms have
been proposed \cite{hammond}. A class of methods try to build the
antisymmetry constraint into the propagator, while
other methods try to reformulate the problem of sampling in the
space of antisymmetric wavefunctions
(determinants) \cite{Shiwei,Hall,Ali}. All these
``fermions'' methods are still at an early stage and their
application has been limited so far to quite small numbers of
fermions. The more robust and widely used, although approximate,
method is the so-called restricted path or fixed node
method \cite{Foulkes01,como95}.

Within the fixed-node method, we need to consider the nodal
surfaces of the fermion density matrix. For any given
configuration $R$, these are defined by the implicit equation
$\rho_{_{F}}(R,R',t)=0$, as the locations $R'$ at which the
density matrix at time $t$ vanishes. The nodal surfaces of the
initial configuration $R$ divide the configurational space of $R'$
in regions of positive $\rho_{_{F}}$ and regions of negative
$\rho_{_{F}}$. In terms of individual paths, the nodal locations
are hypersurfaces in configurational space on which the sum of
contributions of the positive and negative paths to the density
matrix vanishes. Since the fermion density matrix satisfies the
usual convolution relation \cite{FeynmanStatMech}
\begin{equation}
\rho_{_{F}}(R,R',t)=\int dR'' \rho_{_{F}}(R,R'',\tau)\rho_{_{F}}(R'',R',t-\tau)
\hskip 1cm \forall\tau\in[0,t]
\end{equation}
the configurations $R''$ belonging to the nodal surface of the
initial point $R$ at the arbitrary time $\tau$ will not contribute
to the integral and therefore to the density matrix at any future
time $t$. Therefore in constructing the fermion density matrix
from $R$ to $R'$ at time $t$ as sum over signed paths, we can
safely disregard all those paths which have reached the nodal
surface at any previous time $\tau\le t$. If we define the reach
of $R$ at time $t$, $\Upsilon(R,t)$, as the set of points that can
be reached from $R$ in a time $t$ without having crossed the nodal
surfaces at previous times, the argument above can be formalized
in the restricted paths identity\footnote{in ref.
\cite{Ceperley91} an alternative proof based on the Bloch equation
is provided.}
\begin{equation}
\rho_{_{{F}}}(R,R',t)=\frac{1}{N!}\sum_{P}(-)^{P}\left(\int_{Y(0)=R, Y(t)={P}R'} \mathcal{D}Y \hskip 0.1cm e^{-S[Y]}\right)_{\Upsilon(R,t)}
\end{equation}
where $S[Y]$ represent the action of the generic path $Y$ (see eq.
(\ref{eq:DiscreteDM}) for its discretized form). Let us now
consider the generating function $Z(t)$ of eq.
(\ref{eq:GeneratingFunction}). Using the restricted paths identity
we obtain
\begin{eqnarray}
Z(t)&=&\int dR dR' \Psi_{T}(R) \frac{1}{N!}\sum_{P}(-)^{P}\left(\int_{Y(0)=R, Y(t)=PR'} \mathcal{D}Y \hskip 0.1cm e^{-S[Y]}\right)_{\Upsilon(R,t)}  \Psi_{T}(R') \nonumber \\
&=&\int dR dR'' \Psi_{T}(R)\left(\int_{Y(0)=R, Y(t)=R''} \mathcal{D}Y \hskip 0.1cm e^{-S[Y]}\right)_{\Upsilon(R,t)} \frac{1}{N!}\sum_{P}(-)^{P} \Psi_{T}(P^{-1}R'')\nonumber\\
&=&\int dR dR'' \Psi_{T}(R) \left(\int_{Y(0)=R, Y(t)=R''} \mathcal{D}Y \hskip 0.1cm e^{-S[Y]}\right)_{\Upsilon(R,t)}  \Psi_{T}(R'')
\label{eq:RestrictedPathZ(t)}
\end{eqnarray}
where in the last equality we have used the fact that the trial
wave function is antisymmetric under particle permutations:
$(-)^{P}\Psi_{T}(\hat{P}^{-1}R)=\Psi_{T}(R)$. In this last form
the integrand is always positive since for each $R$, the
functional integral is restricted to paths inside its reach so
that $\Psi_{T}(R)\Psi_{T}(R')\ge 0$ (if the nodes of the trial
function are correct). Therefore using the restricted path
identity we have proven that the generating function is a positive
function at any time $t$ and can be computed considering only positive paths which do not cross the nodal surfaces. 

The restricted paths identity is by no means the solution of the
sign problem since in order to know the nodal surfaces we have to
know the density matrix itself. However rephrasing the problem in
terms of spacial boundary conditions can lead to interesting
approximate schemes. The nodal surface of the fermion
density matrix for a system of N interacting particles is a highly
non-trivial function in
6N dimensions and not much is known about it \cite{Ceperley91,Bressanini}. 
The approximate method, known as fixed node in ground state
QMC \cite{Foulkes01} and as restricted paths in PIMC \cite{como95}, consists in replacing the nodal
surfaces of the exact fermion density matrix with the nodal
surfaces of some trial density matrix. In ground state QMC, it is
customary to restrict the class of possible nodal surfaces to time
independent nodes and, within this class, the most reasonable
choice is to assume the nodes of the trial wave function $\Psi_{T}$. 
In practice, this step requires a very minor modification of the
algorithm: it is enough to ensure that
$\Psi_{T}(R_{k-1})\Psi_{T}(R_{k})>0$ for any time interval along
the sampled paths. In the continuous limit this restriction will enforce the restricted path identity.

The nodal surfaces of the trial wave function divide the
configurational space in disconnected regions. In order to perform
the configurational integral over $R$ and $R'$ in eq.
(\ref{eq:RestrictedPathZ(t)}) it could appear necessary to sample
all nodal regions. However, it can be proved that the nodal
regions of the ground state of any Hamiltonian with a reasonable
local potential are all equivalent by symmetry (Tiling
theorem) \cite{Ceperley91}. This ``tiling'' theorem ensures that
computing in a single nodal region is equivalent to a global
calculation.

A further important property of the Fixed node method is the
existence of a variational theorem: the FN-RQMC energy is an upper
bound of the true ground state energy $E_{T}(\infty)\ge E_{0}$,
and the equality holds if the trial nodes coincide with the nodes
of the exact ground state \cite{Foulkes01}. Therefore for fermions,
even projection methods such as RQMC are variational with respect
to the nodal positions; the nodes are not optimized by the
projection mechanism. The ''quality'' of the nodal location is important to obtain accurate results.

In some cases it is necessary to consider complex trial functions,
for instance in the presence of a magnetic field
\cite{OrtizCeperleyMartin93} or in the twist average method to be
discussed later\cite{LinZongCeperley01}. In these cases we have to
deal with a trial function of the form
\begin{equation}
\Psi_{T}(R)=|\Psi_{T}(R)| e^{i\varphi_{_{T}}(R)}
\label{eq:complex-trial}
\end{equation}
where $\varphi_{_{T}}(R)$, a real function, is the configuration
dependent phase of the wave function. Obviously in VMC no
differences arise in the use of complex trial functions other than
in the estimators for the averages. For instance, the local energy
is modified (confr. eq. (27)) and contains an imaginary part
\begin{equation}
E_{L}(R)=V(R)-\lambda\frac{\nabla^{2}|\Psi_{T}|}{|\Psi_{T}|}+\lambda\left(\nabla\varphi_{_{T}}\right)^{2}
-i\lambda\left(2\nabla\varphi_{_{T}}\frac{\nabla|\Psi_{T}|}{|\Psi_{T}|} + \nabla^{2}\varphi_{_{T}}\right)
\label{eq:complex-Eloc}
\end{equation}
Here, we limit the discussion to systems with time-reversal
symmetry, i. e. zero magnetic fields. As $\Psi_{T}$ approaches the
exact ground state, the local energy approaches a real constant
equal to the ground state energy while the imaginary part of the
local energy vanishes. For general complex functions, we can split
the time independent Schr\"odinger equation into two coupled
equations, one for the modulus $|\Phi|$ and one for the phase
$\varphi$ of the wave function
\begin{eqnarray}
\label{eq:SSEmodulo}
&& \left\{-\lambda \nabla^{2} + \left[V(R)+\lambda(\nabla\varphi)^{2}\right]\right\}|\Phi|=E |\Phi| \\
&& \nabla^{2}\varphi + 2 \nabla\varphi\cdot\frac{\nabla|\Phi|}{|\Phi|}=0
\end{eqnarray}
It is therefore natural to take the real part of eq. (\ref{eq:complex-Eloc}) as energy estimator and,
in addition to the variance, to monitor the deviation of the imaginary part from zero as an
indicator of the quality of the trial wave function.\\
How do we have to modify the RQMC to work with complex wave
functions? In the ``fixed phase'' approximation
\cite{OrtizCeperleyMartin93,OrtizCeperley95} one keeps the phase
$\varphi_{_{T}}(R)$ fixed to some analytic form during the
calculation, and solves the imaginary time dependent Schr\"odinger
equation corresponding to the stationary problem of eq.
(\ref{eq:SSEmodulo}). Even for fermions this is a bosonic
problem (since the modulus of the wave function must be symmetric
under particle exchange) with a modified interaction
$\left[V(R)+\lambda(\nabla\varphi)^{2}\right]$. We can still
perform the IS transformation and the formalism remains the same
if the local energy is defined as the real part of eq.
(\ref{eq:complex-Eloc}). Note that the fixed node constraint for
real trial functions can be recast into the fixed phase algorithm
if we write
$\varphi_{T}(R)=\pi[1-\theta(\frac{\Psi_{T}(R)}{|\Psi_{T}(R)|})]$
so that the phase of the trial function changes by $\pi$ across
the nodes at it should. Since the phase is a step function, its
gradient is a $\delta$ function and provides an infinite
contribution in the action of paths crossing the nodes, i.e. a
vanishing probability to cross the nodes.

In this section we have not explicitly indicated the dependence on
the ionic state $S$. In the BO approximation, ions play the role
of external fields for the electronic system so that their
coordinates appear explicitly in the Hamiltonian, in the trial
state, in the local energy and in drift force for the IS
procedure.

\subsection{Trial wave functions for hydrogen}

In this subsection, we describe some general properties of the
trial wave functions for electronic systems. We will restrict our
discussion to the case of a proton-electron system, i. e.
hydrogen and refer to the literature for more complex systems
(heavier elements)\cite{hammond,Foulkes01}. In particular, we will not
discuss the use of pseudopotentials in QMC and the related trial wave functions which, however,
will be an important issue in future extensions of CEIMC.

\subsubsection{The Pair Product Trial Function}
The pair product trial wave function is the simplest extension of the Slater determinant
of single particle orbitals used in mean field treatment of electronic systems (HF or DFT).
This is also the ubiquitous form for trial functions in VMC
\begin{equation}
\Psi_{SJ}(R,\Sigma|S)=exp\left(-\sum_{i<j} u_{ij}(r_{ij})\right) det[\theta_{k}(\vec{r}_{i},\sigma_{i}|S)]
\end{equation}
where $\Sigma=\{\sigma_{1},\ldots,\sigma_{N_{e}}\}$ is the set of spin variables of the electrons, $\theta_{k}(\vec{r},\sigma|S)$ is the $k^{th}$ spin orbital for the given nuclear configuration and
$\theta_{k}(\vec{r}_{i},\sigma_{i}|S)$ is the Slater matrix.
The additional term $u_{ij}(r_{ij})$ is the ``pseudopotential'' or pair correlation factor which introduces explicitly
the two body correlations into the many body wave function. This term is of bosonic nature, i.e. is symmetric
under particle exchange, while the antisymmetry is ensured by the determinant. Often the general form of
both $\theta_{k}$ and $u_{ij}$ are derived by some appropriate theory and then used in connection with some
free variational parameters to be optimized. \\
Let us discuss first the appropriate form of the
``pseudopotential'' for Coulomb systems. There are important
analytical properties for the ``pseudopotential''  that can be
easily derived. Consider bringing two particles together and let
us examine the dominant terms in the local energy. In a good trial
action, the singularities in the kinetic energy must compensate
those in the potential energy. The local energy for the two
particle system is
\begin{equation}
e_{L}(r_{ij})=e^{u_{ij}}\hat{h}_{ij}[e^{-u_{ij}}] = v_{ij}(r_{ij}) + \lambda_{ij} \left[ \nabla^{2}u_{ij}-(\nabla u_{ij})^{2} \right]
\label{eq:2b-Eloc}
\end{equation}
where $\lambda_{ij}=\lambda_{i}+\lambda_{j}$, $\hat{h}_{ij}$ is
the two body hamiltonian with the interaction potential $v_{ij}$
and the trial wave function $\exp[-u_{ij}]$. Spin symmetry has been disregarded as well as the trivial term related to the center of mass motion.
Therefore the short distance behavior of any good form of the
pseudopotential should follow the solution of the two body
Schr\"odinger equation. For the Coulomb potential the ``cusp''
condition derives from this constraint. Indeed substituting $v_{ij}=e^{2}z_{i}z_{j}/r$ in eq.(\ref{eq:2b-Eloc}) and zeroing the coefficient of the dominant power of $r$ for
$r\rightarrow0$ provides
\begin{equation}
\left.\frac{du_{ij}}{dr} \right|_{0}=-\frac{e^{2}z_{i}z_{j}}{\lambda_{ij}(D-1)}
\end{equation}
It is also important to reproduce the correct
behavior at large distances where a description in terms of
collective coordinates is appropriate. The long-wavelength modes
are important for the low energy response properties and are also
the slowest modes to converge in QMC. 
It is possible to show that, within the Random Phase Approximation (RPA), the local energy is minimized by imposing 
\begin{eqnarray}
u_{k}^{ee}&=&-\frac{1}{2}+\sqrt{1+a_{k}} \\
u_{k}^{ep}&=&\frac{-a_{k}}{\sqrt{1+a_{k}}}
\end{eqnarray}
with $a_{k}=12 r_{s}/k^{4}$ and the electron sphere radius $r_{s}$ is related to the volume per electron by $v=4\pi r_{s}^{3}/3$ in atomic units \cite{natoli-theses}. The obtained form of the ``pseudopotential'' is correct at short
and long distances but not necessarily in between because of the
approximation. One can improve slightly the quality of the VMC
results considering the form
\begin{equation}
\tilde{u}_{ij}(r)=u^{RPA}_{ij}(r)-\alpha_{ij}e^{-{r^{2}}/{w_{ij}^{2}}}
\end{equation}
with the variational parameter $\alpha_{ij}, w_{ij}$. The
additional term preserves the short and long distance behavior of
the RPA function. This form of the pair trial function introduced
four variational parameters, namely $\alpha_{ee}, w_{ee},
\alpha_{ep}, w_{ep}$.

We discuss now the choice of the spin orbitals. The spin-orbitals
are conceptually more important than the pseudopotential because
they provide the nodal structure of the trial function. With the
fixed node approximation, the projected ground state has the same
nodal surfaces of the trial function, while the other details of
the trial function are automatically ``optimized'' for increasing
projection time. It is thus important that the nodes provided by
given spin-orbitals be accurate. Moreover, the optimization of
nodal parameters  (see below) is, in general, more difficult and
unstable than for the pseudopotential parameters \cite{Foulkes01}.

The simplest form of spin-orbitals for a system with translational
invariance are plane waves (PW)
$\theta_{\vec{k}}(\vec{r},\sigma)=exp[\imath
\vec{k}\cdot\vec{r}]$. This form was used in the first QMC study
of metallic hydrogen \cite{CeperleyAlder87}. It is particularly
appealing for its simplicity and still qualitatively correct since
electron-electron and electron-proton correlations are considered
through the ``pseudopotential''. The plane waves orbitals are
expected to reasonably describe the nodal structure for metallic
atomic hydrogen, but no information about the presence of protons
appears in the nodes with PW orbitals.

For insulating molecular hydrogen (i.e for $r_{s}\geq 1.5$), it is
preferred to use localized gaussian orbitals. There are different
possibilities: a single isotropic gaussian centered at the middle
of the bond was used in the first QMC study\cite{CeperleyAlder87},
while a single multivariate gaussian was used in the first CEIMC
attempt\cite{DewingCeperley01,dmc03}. Another possibility is to form a molecular
orbitals as linear combination of two atomic gaussians orbitals
centered on each proton:
\begin{equation}
\theta_{k}(\vec{r},\vec{s}_{k,1},\vec{s}_{k,2})=exp(-c|\vec{r}-\vec{s}_{k,1}|^{2})+
exp(-c|\vec{r}-\vec{s}_{k,2}|^{2})
\label{eq:MolecularOrbitals}
\end{equation}
where $\vec{s}_{k,j}$ is the position of the $j^{th}$ proton of
the $k^{th}$ molecule. This kind of orbitals have a single
variational parameter $c$. At present we are experimenting using a
trial molecular wave function with these orbitals multiplied by
the corrected RPA Jastrow within the CEIMC.

The trouble with this strategy is that one should know which phase
is stable before performing the calculation. This is typical of
ground state studies. However in CEIMC we would rather let the
system find its own state for given temperature and density (or
pressure). In particular, this approach is not appropriate to
address the interesting region of molecular dissociation and
metallization. This problem can be solved by using orbitals
obtained as solution of a single-electron problem as in band
structure calculations or in self-consistent mean field methods.
In previous works on ground state hydrogen, the single electron
orbitals for a given protonic state S, were obtained from a
DFT-LDA calculation \cite{WangZhuLouieFahy90,natoli93,natoli95}. One of this study \cite{natoli93} established
that energies from plane-waves determinants in metallic hydrogen
are higher than the more accurate estimates from DFT-LDA orbitals
by 0.05eV/atom at the density at which the transition between
molecular and metallic hydrogen is expected ($r_{s}=1.31$).
Obtaining the orbitals from a DFT-LDA calculation has, however,
several drawbacks in connection with the CEIMC. While for protons
on a lattice we can solve the self-consistent theory for a
primitive cell only, in a disordered configuration, we need to
consider the entire simulation box. This is very expensive in CPU
time and memory for large systems. Moreover, combining the LDA
orbitals with Jastrow to improve the accuracy is not
straightforward; substantial modification of the orbitals might be
necessary requiring a reoptimization of the orbitals and the
correlation factors, in principle, at each new ionic
position. 

\subsubsection {Beyond the pair product trial action}
Over the years there have been important progress in finding trial
functions substantially more accurate then the pair product form
for homogeneous systems \cite{panoffcarlson,kwon}. Within the
generalized Feynman-Kac formalism, it is possible to systematically
improve a given trial function\cite{kwon,hcpe03}. The first
corrections to the pair product action with plane wave orbitals
are a three-body correlation term which modifies the correlation
part of the trial function (Jastrow) and a ``backflow''
transformation which changes the orbitals and therefore the nodal
structure (or the phase) of the trial function\cite{hcpe03}. The
new trial function has the form
\begin{equation}
\Psi_{T}(R,\Sigma|S)=det[\theta_{k}(\vec{x}_{i},\sigma_{i}|S)]e^{-U_{2}-U_{3}}
\end{equation}
where $U_{2}=\sum_{i<j}\tilde{u}_{ij}$ is the two body ``pseudopotential'' discussed before, $U_{3}$ the three-body term of the form
\begin{equation}
U_{3}=-\sum_{i=1}^{N_{e}}\left[\sum_{j=1}^{N}\xi_{ij}(r_{ij})\vec{r}_{ij}\right]^{2}
\label{eq:3b}
\end{equation}
and finally the ``quasiparticle'' coordinates appearing in the plane wave orbitals are given by
\begin{equation}
\vec{x}_{i}=\vec{r}_{i}+\sum_{j=1}^{N}\eta_{ij}(r_{ij})\vec{r}_{ij}; \hskip 0.5cm (i=1,\cdots,N_{e})
\label{eq:BF}
\end{equation}
The functional form of the three-body term is that of a squared
two-body force so its evaluation is not slower than a genuine
2-body term. In the homogeneous electron system, this term is
particularly relevant at low density where correlation effects are
dominant. On the other hand, the backflow transformation is more
relevant at high density because in this limit the fermionic
character of the system dominates. The same general framework
should hold for hydrogen although a throughout study of the
relative importance of those effects with density is still
missing. The fundamental improvement of backflow orbitals for
metallic hydrogen is that the nodal structure of the wave
functions depends now on the proton positions. This provides
better total energies for static protonic
configurations \cite{hcpe03} and improved energy differences and
liquid structure in CEIMC \cite{PC05,dmc03}.
However it is not clear how appropriate this kind of wave function
will be when entering in the molecular phase of hydrogen.

The unknown functions, $\xi_{ij}(r), \eta_{ij}(r)$ in eqs.
(\ref{eq:3b}) and (\ref{eq:BF}) need to be parameterized in some
way.  In a first attempt we have chosen gaussians with variance
and amplitude as new variational parameters \cite{dmc03}. This form
was shown to be suitable for homogeneous electron gas \cite{kwon}.
Approximate analytical forms for $\xi_{ij}(r)$ and $\eta_{ij}(r)$,
as well as for the two-body pseudopotential, have been obtained
later in the framework of the Bohm-Pines collective coordinates
approach \cite{hcpe03}. This form is particularly suitable for the
CEIMC because there are no parameters to be optimized. This trial
function is faster than the pair product trial function with the
LDA orbitals, has no problems when protons move around and its
nodal structure has the same quality as the corresponding one for the LDA Slater
determinant \cite{hcpe03}. We have extensively used this form of
the trial wave function for CEIMC calculations of metallic atomic
hydrogen.

\subsubsection{Trial function optimization}
For metallic hydrogen we have described a parameter-free trial
function which does not need optimization. However, if we use the
pair proton action both for molecular or LDA orbitals, we are left
with free parameters in the Jastrow factor and with the width of
the gaussians for molecular orbitals. 
Optimization of the parameters in a trial function is crucial for
the success of VMC. Bad upper bounds do not give much physical
information. Good trial functions will be needed in the Projector
Monte Carlo method. First, we must decide on what to optimize and
then how to perform the optimization. There are several
possibilities for the quantity to optimize and depending on the
physical system, one or other of the criteria may be best.

\begin{itemize}
\item The variational energy: $E_V$. If the object of the
calculation is to find the least upper bound one should minimize
$E_V$. There is a general argument suggesting that the trial
function with the lowest variational energy will maximize the
efficiency of Projector Monte Carlo \cite{c37}. 
\item The variance
of the local energy: $\sigma^2 =  \int |{\mathcal H}\Psi|^2
-E_V^2$ . If we assume that every step on a QMC calculation is
statistically uncorrelated with the others, then the variance of
the average energy will equal $\sigma^2/p$ where $p$ is the number
of steps. The minimization of $\sigma^2$ is statistically more
robust than the variational energy because it is a positive
definite quantity with zero as minimum value. One can also
minimize a linear combination of the variance and the variational
energy. \item The overlap with the exact wave function: $\int \Psi
\phi$. If we maximize the overlap, we find the trial function
closest to the exact wave function in the least squares sense.
This is the preferred quantity to optimize if you want to
calculate correlation functions, not just ground state energies
since, then, the VMC correlation functions will be closest to the
true correlation functions. Optimization of the overlap will
involve a Projector Monte Carlo calculation to determine the
change of the overlap with respect to the trial function so it is
more complicated and rarely used.
\end{itemize}
The most direct optimization
method consists of running independent VMC calculations using
different set of numerical values for the variational parameters.
One can fit the energies to a polynomial, performing more
calculations near the predicted minimum and iterating until
convergence in parameter space is attained.  The difficulty  with
this direct approach is that close to the minimum, the independent
statistical errors will mask the variation with respect to the
trial function parameters. This is because the derivative of the
energy with respect to trial function parameters is very poorly
calculated. Also, it is difficult to optimize, in this way,
functions involving more than 3 variational parameters because so
many independent runs are
needed to cover the parameter space.\\
A correlated sampling method, known as reweighting \cite{c9,c3} is
much more efficient. One samples a set of configurations $\{ R_j
\}$ (usually several thousand points at least) according to some
distribution function, usually taken to be the square of the
wavefunction for some initial trial function: $|\Psi_{T}(R;
\{\alpha\}_0)|^2$. Then, the variational energy (or the variance)
for trial function nearby in parameter space can be calculated by
using the same set of points:
\begin{equation}
E_v (a) = \frac{ \sum_j w(R_j, \{\alpha\}) E_L (R_j, \{\alpha\}) }
{ \sum_j w(R_j,\{\alpha\}) },
\end{equation}
where  the weight factor, $w(R) = | \Psi_T (R ; \{\alpha\})/\Psi_T
(R; \{\alpha\}_0) |^2$, takes into account that the distribution
function changes as the variational parameters change.  One then
can use a minimizer to find the lowest variational energy or
variance as a function of $\{\alpha\}$ keeping the configurations
fixed. However, there is an instability: if the parameters move
too far away, the weights span too large of a range and the error
bars of the energy become large. The number of effective points of
a weighted sum is:
\begin{equation}
N_{eff} = (\sum w_j)^2/ \sum w_j^2 .
\end{equation}
If this becomes much smaller than the number of points, one
must resample and generate some new points. When minimizing the variance,
one can also simply neglect the weight factors.
Using the reweighting method one can find the optimal value
of wavefunction containing tens of parameters.

\subsection{Twist Average Boundary Conditions}

Almost all QMC calculations in periodic
boundary conditions have assumed that the phase of the wave
function returns to the same value if a particle goes around the
periodic boundaries and returns to its original position. However,
with these boundary conditions, delocalized fermion systems
converge slowly to the thermodynamic limit because of shell
effects in the filling of single particle states. Indeed, with periodic boundary conditions
the Fermi surface of a metal will be reduced to a discrete set of points in k-space.
The number of k-points is equal to the number of electrons of same spin and therefore it is quite limited. \\
One can allow particles to pick up a phase when they wrap around the periodic boundaries,
\begin{equation}
\Psi_{\vec{\theta}}(\vec{r}_1 + L  \vec{n}, \vec{r}_2,
\cdots )= e^{i\vec{\theta}} \Psi_{\vec{\theta}}(\vec{r}_1,\vec{r}_2, \cdots ).
\label{eq:TBC}
\end{equation}
where we have assumed a cubic box of size $L$ and $\vec{n}$ is a
vector of integers. The boundary condition $\vec{\theta} = 0$ is
periodic boundary conditions (PBC), and the general condition with
$\vec{\theta} \neq 0$, twisted boundary conditions (TBC). If the
periodic boundaries are used in all directions, each dimension can
have an independent twist. Hence, in three dimension (3D), the
twist angle is a three component vector. The free energy and
therefore all equilibrium properties are periodic in the twist:
$F(\vec{\theta}+2\pi \vec{n})=F(\vec{\theta})$ so that each
component of the twist
can be restricted to be in the range $-\pi<\theta_{i}\leq\pi$.\\
The use of twisted
boundary conditions is commonplace for the solution of the band
structure problem for a periodic solid, particularly for metals.
In order to calculate properties of an infinite periodic solid,
properties must be averaged by integrating over the first
Brillouin zone.

For a degenerate Fermi liquid, finite-size shell effects are much
reduced if the twist angle is averaged over: twist averaged
boundary conditions (TABC)\cite{LinZongCeperley01}. For any given property $\hat{A}$ the TABC is defined as
\begin{equation}
<\hat{A}>=\int_{-\pi}^{\pi}\frac{d\vec{\theta}}{(2\pi)^{d}} <\Psi_{\vec{\theta}}|\hat{A}|\Psi_{\vec{\theta}}>
\end{equation}
TABC is particularly important in computing properties that are
sensitive to the single particle energies such as the kinetic
energy and the magnetic susceptibility. By reducing shell effects,
accurate estimations of the thermodynamic limit for these
properties can be obtained already with a limited number of
electrons. What makes this very important is that the most
accurate quantum methods have computational demands which increase
rapidly with the number of fermions. Examples of such methods are
exact diagonalization  (exponential increase in CPU time with N),
variational Monte Carlo (VMC) with wave functions having backflow
and three-body terms \cite{kwon} (increases as $N^4$),
and transient-estimate and released-node Diffusion Monte Carlo
methods \cite{cep84} (exponential increase with N). Moreover, size
extrapolation is impractical within CEIMC since it would have to
be performed for any proposed ionic move prior to the acceptance
test. Methods which can extrapolate more rapidly to the
thermodynamic limit are crucial in obtaining high accuracy.

Twist averaging is especially advantageous in combination with
stochastic methods (i.e.  QMC) because the twist averaging does not necessarily slow
down the evaluation of averages, except for the necessity of doing
complex rather than real arithmetic. In a metallic system, such as
hydrogen at very high pressure,
results in the thermodynamic limit require careful integration
near the Fermi surface because the occupation of states becomes
discontinuous. Within LDA this requires ``k--point'' integration,
which slows down the calculation linearly in the number of
k-points required. Within QMC such k-point integration takes the
form of an average over the (phase) twist of the boundary
condition and can be done in parallel with the average over
electronic configurations without significantly adding to the
computational effort.

In CEIMC we can take advantage of the twist averaging to reduce
the noise in the energy difference for the acceptance test of the
penalty method (see below). In the electron gas, typically 1000
different twist angles are required to achieve
convergence \cite{LinZongCeperley01}. We have used the same number
of twist angles in CEIMC calculations of metallic hydrogen.
Different strategies can be used to implement the
TABC \cite{LinZongCeperley01}. We have used a fixed 3D grid in the
twist angle space, at each grid point run independent QMC
calculations and then averaged the resulting properties. This
procedure can be easily and efficiently implemented on a parallel
computer. Recently, we have devised a sampling procedure to
randomize the grid points at each ionic step. We have limited
experience, but, so far, we have evidence that good convergence in
the electronic and ionic properties can already be reached for a
number of twist angles as low as 30 \cite{PDCtbp}.

\subsection{Sampling electronic states: the ``bounce'' algorithm}
In this section we describe the way we have implemented electronic
move in the CEIMC method. In particular we present an original
algorithm for RQMC, particularly suitable for CEIMC, called the
``bounce'' algorithm.

First, how do the particles move in VMC? In the continuum it
is usually more efficient to move the particles one at a time by
adding a random vector to a particle's coordinate, where the
vector is either uniform inside of a cube, or is a normally
distributed random vector centered around the old position.
Unfortunately, this procedure cannot be used with backflow
orbitals. This is because the backflow transformation couples all
the electronic coordinates in the orbitals so that once a single
electron move is attempted the entire Slater determinant needs to
be recomputed, an $O(N^{3})$ operation. It is much more efficient
to move all electrons at once, although global moves could become
inefficient for large systems.

Next we describe the electronic sampling within RQMC.
In the original work on RQMC \cite{BaroniMoroni}, the electronic
path space was sampled by a simple reptation algorithm, an algorithm
introduced to sample the configurational space of linear polymer
chains \cite{FrenkelSmit}.
Remember that in RQMC the electronic configurational space is the
space of $3N_{e}$-dimensional random paths of length ``t''. In
practice, the imaginary time is discretized in $M$ time slices
$\tau=t/M$ and the paths become discrete linear chains of $M+1$
beads. Let us indicate with $Q$ the entire set of $3N(M+1)$
coordinates $Q=\{R_{0},\ldots,R_{M}\}$. According to eqs.
(\ref{eq:linkaction}), (\ref{eq:DiscreteDM}) and
(\ref{eq:ZetaRQMC}), the path distribution is
\begin{equation}
\Pi(Q)=|\Psi_{T}(R_{0})\Psi_{T}(R_{M})|e^{-\sum_{k=1}^{M}L_{s}(R_{k-1},R_{k},\tau)}
e^{-\tau\left[\frac{E_{L}(R_{0})}{2}+\sum_{k=1}^{M-1}E_{L}(R_{k})+\frac{E_{L}(R_{M})}{2}\right]}
\end{equation}
Given a path configuration $Q$, a move is done in two stages.
First one of the two ends (either $R_{0}$ or $R_{M}$) is sampled
with probability 1/2 to be the growth end $R_{g}$. Then a new
point near the growth end is sampled from a Gaussian distribution
with center at $R_{g}+2\lambda\tau_{e}F(R_{g})$. In order to keep
the number of links on the path constant, the old tail position is
discarded in the trial move. The move is accepted or rejected with
the Metropolis formula based on the probability of a reverse move.
For use in the following, let us define the direction variable $d$
as $d=+1$ for a head move ($R_{g}=R_{M}$), and $d=-1$ for a tail
move ($R_{g}=R_{0}$). In standard reptation, the direction $d$ is
chosen randomly at each attempted step. The transition probability
$P(Q\rightarrow Q')$ is the product of an attempt probability
$T_{d}(Q\rightarrow Q')$ and an acceptance probability
$a_d(Q\rightarrow Q')$. The paths distribution $\Pi(Q)$ does not
depend on the direction $d$ in which it was constructed. In the
Metropolis algorithm, the acceptance probability for the attempted
move is
\begin{equation}
a_d(Q\rightarrow Q')=min\left[1,\frac{\Pi(Q')T_{-d}(Q'\rightarrow
Q)}{\Pi(Q)T_{d}(Q\rightarrow
Q')}\right]\label{eq:acceptance}
\end{equation}
which ensures that
the transition probability $P_d(Q \rightarrow Q')$ satisfies
detailed balance
\begin{equation}
\Pi(Q)P_d(Q\rightarrow Q')=\Pi(Q')P_{-d}(Q'\rightarrow Q)
\label{eq:DB}
\end{equation}

The autocorrelation time of this algorithm, that is the number of
MC steps between two uncorrelated configurations, scales as
$[M^{2}/A]$, where $A$ is the acceptance rate of path moves, an
unfavorable scaling for large $M$ (i.e. large projection time
$t$). Moreover the occasional appearance of persistent
configurations bouncing back and forth without really sampling the
configuration space has been previously observed \cite{saverio}.
These are two very unfavorable features, particularly in CEIMC,
where we need to perform many different electronic calculations.
There is a premium for a reliable, efficient and robust algorithm.

We have found that a minimal modification of the reptation
algorithm solves both of these problems. The idea is to chose
randomly the growth direction at the beginning of the Markov
chain, and reverse the direction upon rejection only, the
``bounce'' algorithm. 

What follows is the proof that the bounce algorithm samples the
correct probability distribution $\Pi(Q)$. The variable $d$ is no
longer randomly sampled, but, as before, the appropriate move is
sampled from the same Gaussian distribution $T_{d}(Q\rightarrow
Q')$ and accepted according to the Eq. (\ref{eq:acceptance}). In order
to use the techniques of Markov chains, we need to enlarge
the state space with the direction variable $d$. In the enlarged
configuration space $\left\{Q,d\right\} $, let us define the
transition probability $P(Q,d\rightarrow Q',d')$ of the Markov
chain. The algorithm is a Markov process in the extended path
space, and assuming it is ergodic, it must converge to a unique
stationary state, $\Upsilon(Q,d)$ satisfying the eigenvalue
equation:
\begin{equation}
\sum_{Q,d}\Upsilon(Q,d)\, P(Q,d\rightarrow Q',d')=\Upsilon(Q',d')
\label{eq:eigenvalue}.
\end{equation}
We show that our desired probability $\Pi(Q)$ is solution of this
equation. Within the imposed rule, not all transitions are
allowed, but $P(Q,d\rightarrow Q',d')\neq 0$ for $d=d'$ and $Q\neq
Q'$ (accepted move), or $d'=-d$ and $Q=Q'$ (rejected move) only.
Without loss of generality let us assume $d'=+1$ since we have
symmetry between $\pm1$. Eq. (\ref{eq:eigenvalue}) with
$\Upsilon(Q,d)$ replaced by $\Pi(Q)$ is
\begin{equation}
\Pi(Q')P(Q',-1\rightarrow Q',1)+\sum_{Q\neq
Q'}\Pi(Q)P(Q,1\rightarrow Q',1)=\Pi(Q').
\end{equation}
Because of detailed
balance Eq.(\ref{eq:DB}), we have
$$\Pi(Q)P(Q,1\rightarrow Q',1)=\Pi(Q')P(Q',-1\rightarrow Q,-1)$$
which, when substituted in this equation gives
\begin{equation}
\Pi(Q')\left[P(Q',-1\rightarrow Q',1)+\sum_{Q}P(Q',-1\rightarrow Q,-1)\right]=\Pi(Q').
\end{equation}
Note that we have completed the sum over $Q$ with the term $Q=Q'$ because
its probability vanishes. The term in the bracket exhausts all
possibilities for a move from the state $(Q',-1)$, thus it adds to
one. Hence $\Pi(Q)$ is a solution of eq. (\ref{eq:eigenvalue}) and
by the theory of Markov chains, it is the probability distribution
of the stationary state.

\section{Sampling ionic states: The penalty method}
In Metropolis Monte Carlo a Markov chain of ionic states $S$ is
generated according to the Boltzmann distribution $P(S) \propto
e^{-\beta E_{BO}(S)}$ where $E_{BO}(S)$ is the Born-Oppenheimer
energy for the ionic configuration $S$ and $\beta$ the inverse
temperature. From the state $S$ a trial state $S'$ is proposed
with probability $T(S\rightarrow S')$ and the detailed balance
condition is imposed by accepting the move with probability
\begin{equation}
A(S\rightarrow S')=min\left[1,\frac{T(S'\rightarrow S) e^{-\beta E_{_{BO}}(S')}}{T(S\rightarrow S') e^{-\beta E_{_{BO}}(S)}}\right]
\end{equation}
Under quite general conditions on the system and on the a-priori transition probability $T(S\rightarrow S')$,
after a finite number of MC steps the Markov chain so generated will visit the states of the
configurational space with a frequency proportional to their Boltzmann's weight \cite{FrenkelSmit}.\\
In CEIMC estimate of $E_{BO}(S)$ is affected by statistical noise.
If we ignore the presence of noise and we use the standard
Metropolis algorithm, the results will be biased, the amount of
bias increasing for increasing noise level. A possible solution
would be to run very long QMC calculations in order to get a
negligibly small noise level resulting in a negligible bias.
However, the noise level decreases as the number of independent
samples to the power 1/2, that is to decrease the noise level by
one order of magnitude we should run 100 times longer, an
unfavorable scaling if we realize that we have to repeat such
calculation for any attempted move of the ions. The less obvious
but far more efficient solution is to generalize the Metropolis
algorithm to noisy energies. This is done by the Penalty
Method\cite{CeperleyDewing99}. The idea is to require the detailed
balance to hold on average and not for any single energy
calculation.

Let us consider two ionic states $(S, S')$ and call $\delta(S,S')$
the ``instantaneous'' energy difference times the inverse temperature.
Let us further assume that the average and the variance of
$\delta(S,S')$ over the noise distribution $P(\delta|S\rightarrow
S')$ exhist
\begin{eqnarray}
\Delta(S,S')&=&\beta\left]E_{BO}(S')-E_{BO}(S)\right]=<\delta(S,S')> \\
\sigma^{2}(S,S')&=&<(\delta(S,S')-\Delta(S,S'))^{2}>
\end{eqnarray}
We introduce the ``instantaneous'' acceptance probability,
$a(\delta|S,S')$ and impose the detailed balance to hold for the
average of $a(\delta|S,S')$ over the distribution of the noise
\begin{equation}
A(S\rightarrow S')=e^{-\Gamma(S,S')}A(S'\rightarrow S)
\label{eq:DetaliedBalanceOnAverage}
\end{equation}
where
\begin{equation}
A(S\rightarrow S')=\int_{-\infty}^{\infty} d\delta P(\delta|S\rightarrow S') a(\delta|S,S')
\label{eq:AverageAcceptance}
\end{equation}
and we have defined
\begin{equation}
\Gamma(S,S')=\Delta(S,S') - \ln\left[\frac{T(S'\rightarrow S)}{T(S\rightarrow S')}\right]
\label{eq:Gamma(S,S')}
\end{equation}
If we assume the quite general conditions
$a(\delta|S,S')=a(\delta)$ and $P(\delta|S\rightarrow S')=P(-\delta|S'\rightarrow S)$
to hold, the detailed balance can be written
\begin{equation}
\int_{-\infty}^{\infty}d\delta P(\delta|S\rightarrow S') \left[a(\delta)-e^{-\Gamma}a(-\delta)\right]=0
\label{eq:DBintegral}
\end{equation}
which, supplemented by the condition $a(\delta)\geq0$, is the
equation to solve in order to obtain the acceptance probability
$a(\delta)$. The difficulty is that during the MC calculation we
do not know either
$P(\delta|S\rightarrow S')$ nor $\Delta(S,S')$. \\
In order to make progress let us assume, as it happens in many
interesting cases, that the noise of the energy difference is
normally distributed so that
\begin{equation}
P(\delta|S\rightarrow S')=(2\pi\sigma^{2})^{-1/2} \exp\left[-(\delta-\Delta)^{2}/2\sigma^{2}\right]
\end{equation}
Let us, moreover, assume that we know the value of the variance
$\sigma$. It is not difficult to check that the solution of eq.
(\ref{eq:DBintegral}) is
\begin{equation}
a_{n}(\delta|\sigma)=min\left[1,\frac{T(S'\rightarrow S)}{T(S\rightarrow S')}\exp\left(-\delta-\frac{\sigma^{2}}{2}\right)\right]
\label{eq:GeneralizedAcceptance}
\end{equation}
The uncertainty in the energy difference just causes a reduction
in the acceptance probability by an amount $\exp(-\sigma^{2}/2)$
for $\delta>-\sigma^{2}/2$. The integral of $a_{n}(\delta|\sigma)$
over the guassian measure provides
\begin{eqnarray}
A(S\rightarrow S')=&\frac{1}{2}&
\mathrm{erfc}\left\{\left[\sigma^{2}/2+\Gamma(S\rightarrow S')\right]/(\sqrt{2}\sigma)\right\} +\nonumber \\
 &\frac{1}{2}&\mathrm{erfc}\left\{\left[\sigma^{2}/2-\Gamma(S\rightarrow S')\right]/(\sqrt{2}\sigma)\right
 \}e^{-\Gamma(S\rightarrow S')}
\end{eqnarray}
which satisfies eq. (\ref{eq:DetaliedBalanceOnAverage}) since
$\Gamma(S'\rightarrow S)=-\Gamma(S\rightarrow S')$. Note that eq.
(\ref{eq:GeneralizedAcceptance}) reduces to the standard
Metropolis form for vanishing $\sigma$ \cite{FrenkelSmit}.

An important issue is to verify that the energy differences are
normally distributed. Recall that if the moments of the energy difference are
bounded, the central limit theorem implies that given enough
samples, the distribution of the mean value will be Gaussian.
Careful attention to the trial function to ensure that the local
energies are well behaved may be needed.

In practice not only the energy difference $\Delta$ but also the
variance $\sigma$ is unknown and must be estimated from the data.
Let us assume that, for a given pair of ionic states $(S,S')$, we
generate $n$ statistically uncorrelated estimates of the energy
difference $\{y_{1},\ldots,y_{n}\}$ each normally distributed with
their first and second moments defined in the usual way
\begin{eqnarray}
\Delta&=&<y_{i}> \\
\sigma^{2}&=&<(y_{i}-\Delta)^{2}>
\end{eqnarray}
Unbiased estimates of $\Delta$ and $\sigma^{2}$ are
\begin{eqnarray}
\delta&=&\frac{1}{n}\sum_{i=1}^{n}y_{i} \\
\chi^{2}&=&\frac{1}{n(n-1)}\sum_{i=1}^{n}(y_{i}-\delta)^{2}
\end{eqnarray}
An extension of the derivation above provides \cite{CeperleyDewing99}
\begin{equation}
a(\delta,\chi^2,n) = \min\left[1,\frac{T(S'\rightarrow S)}{T(S\rightarrow S')}\exp(-\delta-u_B)\right]
\label{eq:BesselAcceptance}
\end{equation}
where
\begin{equation}
  u_B  =\frac{\chi^2}{2} + \frac{\chi^4}{4(n+1)} +\frac{\chi^6}{3(n+1)(n+3)}  + \cdots
\label{eq:BesselExpansion}
\end{equation}
The first term in eq. (\ref{eq:BesselExpansion}) is the penalty in
the case we know the variance. The error of the noise causes extra
penalty which decreases as the number of independent samples $n$
grows. In the limit of large $n$ the first term dominates and we
recover eq. (\ref{eq:GeneralizedAcceptance}). Eq.
(\ref{eq:BesselExpansion}) must be supplemented by the condition
$\chi^{2}/n\leq1/4$ for the asymptotic expansion
(\ref{eq:BesselExpansion}) to converge and the instantaneous
acceptance probability to be positive \cite{CeperleyDewing99}.

The noise level of a system can be characterized by the relative
noise parameter, $f= (\beta \sigma)^2 t/ t_0$, where $t$ is the
computer time spent reducing the noise, and $t_0$ is the computer
time spent on other pursuits, such as optimizing the VMC wave
function or equilibrating the RQMC runs.  A small $f$ means little
time is being spent on reducing noise, where a large $f$ means
much time is being spent reducing noise. For a double well
potential, the noise level that gives the maximum
efficiency is around $\beta \sigma \approx 1$, with the optimal
noise level increasing as the relative noise parameter increases \cite{CeperleyDewing99}.
In CEIMC runs for hydrogen the noise level $\beta \sigma$ ranges between 0.3 and 3, the optimal value being around 1.

\subsection{Efficient strategies for electronic energy differences}
As explained above, we need to evaluate the energy difference and
the noise between two protonic configurations $(S,S')$. The
distance in configurational space between $S$ and $S'$ in an
attempted move is however quite limited since we have to move all
protons at once. Indeed, each backflow orbital depends on the
position of all protons and single proton moves would require
recomputing the entire determinant for each attempted move, a
O($N^{4}$) operations for a global move. Instead, moving all
protons together requires a single determinant calculation per
ionic move, a O($N^{3}$) operation. In this case, performing two
independent electronic calculations for $S$ and $S'$ to estimate
the energy difference would be very inefficient since $\Delta
E_{BO} << E_{BO}(S)$ and the noise on the energy difference would
just be twice the noise on the single energy estimate. The
strategy to adopt is to compute the energy difference from
correlated sampling, i.e. sampling the electronic configurational
space from a distribution which depends both on $S$ and $S'$ and
estimating the energy difference and the other electronic
properties of interest by reweighting the averages \cite{dmc03}. It
is possible to show that the optimal sampling function, i.e. the
sampling distribution for which the variance of the energy
difference is minimal, takes the form
\begin{equation}
P(Q|S,S')\propto \left| \Pi(Q|S)(E_{BO}(S)-<E_{BO}(S)>)-\Pi(Q|S')(E_{BO}(S')-<E_{BO}(S')>)\right|
\end{equation}
where $<E_{BO}>$ is the estimate of the BO energy. In
order to use this sampling probability we need to estimate the BO
energies of the two states before performing the sampling. A
simpler form which avoid this problem is
\begin{equation}
P(Q|S,S')\propto \Pi(Q|S) + \Pi(Q|S')
\end{equation}
We emphasize that the ``reptile'' space for the electron paths
depends on the proton coordinates so that, because of the fixed-node restriction
for fermions, legal paths for $S$ may
or may not be legal for $S'$ and vice-versa. These two forms of
importance sampling have the property that they sample regions of
both configuration spaces (S and S') and make the energy
difference rigorously correct with a bounded variance.

\subsection{Pre-rejection}
We can use multi--level sampling to make CEIMC more efficient
\cite{rmp95}. An empirical potential is used to
``pre-reject'' moves that would cause particles to overlap and be
rejected anyway. A trial move is proposed and accepted or rejected
based on a classical potential
\begin{equation}
A_1 = \min \left[1,\frac{T(S\rightarrow S')}{T(S' \rightarrow S)}
      \exp(-\beta \Delta V_{cl})\right]
\end{equation}
where $\Delta V_{cl} = V_{cl}(S') - V_{cl}(S)$ and $T$ is the
sampling probability for a move. If it is accepted at this first
level, the QMC energy difference is computed and accepted with
probability
\begin{equation}
A_2 = \min \left[1, \exp(-\beta \Delta E_{BO}-u_B ) \exp(\beta \Delta V_{cl}) \right]
\end{equation}
where $u_B$ is the noise penalty. 
Compared to the cost of evaluating the QMC energy difference,
computing the classical energy difference is much less expensive.
Reducing the number of QMC energy difference evaluations reduces
the overall computer time required.

For metallic hydrogen a single yukawa potential  was always found
to be suitable for pre-rejection. For molecular hydrogen we use
instead a Silvera-Goldman potential \cite{silvera78} riparametrized
 in such a way to have the center of
interaction on any single proton in the molecule rather than on
the molecular center of mass. In practice, we take a single Yukawa
potential for intermolecular proton-proton interaction and the 
Kolos-Wolniewski potential \cite{kolos64} for the bonding interaction. 
The Yukawa screening length and the
prefactor are optimized to reproduce the results of the
Silvera-Goldman model. This new potential is suitable for
pre-rejecting all types of moves that we attempt in the molecular
hydrogen, namely molecular rotations, bond stretching and
molecular translations. The original Silvera-Goldmann potential
being spherically symmetric around the molecular center is not
suitable to pre-reject the rotational moves.

\section{Quantum protons}
By increasing pressure and/or decreasing temperature, ionic
quantum effects can become relevant. Those effects are important
for hydrogen at high pressure\cite{maksimivshilov,BabaevSudbeAshcroft04}. 
Static properties of quantum
systems at finite temperature can be obtained with the Path
Integral Monte Carlo method (PIMC) \cite{rmp95}. We need to
consider the ionic thermal density matrix rather than the
classical Boltzmann distribution:
\begin{equation}
\rho_{p}(S,S'|\beta)=<S|e^{-\beta(\hat{K}_p+\hat{E}_{BO})}|S'>
\end{equation}
where $\hat{K}_p$ is the ionic kinetic energy operator and $\beta=(K_{B}T)^{-1}$ is the inverse physical temperature.
Thermal averages of ionic operators $\hat{A}_{p}$ (diagonal in configurational space) are obtained as
\begin{equation}
A_{p}(\beta)=\frac{1}{Z(\beta)}\int dS A_{p}(S) \rho_{p}(S,S|\beta)
\end{equation}
where $Z$ is the partition function
\begin{equation}
Z(\beta)=\int dS \rho_{p}(S,S|\beta)= e^{-\beta F}
\end{equation}
and $F$ is the Helmholtz free energy of the system. As before, the
thermal density matrix can be computed by a factorization of
$\beta$ in many (P) small intervals (time slices
$\tau_{p}=\beta/P$) and by a suitable approximation for the ``high
temperature'' (or short time) density matrix. According to the
Feynman-Kac formula (see eqs. (\ref{eq:FeynmanKac}) and
(\ref{eq:GeneralizeFeynmann-Kac})) the diagonal part of the
thermal density matrix is the sum over all closed paths, i.e.
paths starting at $S$ and returning to $S$ after a ``time''
$\beta$. This is the famous ``isomorphism'' between quantum
particles and ring polymers \cite{FeynmanStatMech,rmp95}. Considering particle
statistics in the PIMC is more difficult than in RQMC. The reason
is that in PIMC the state of the classical system, which has no
symmetry built in, plays the role of the trial functions in RQMC
and permutations need to be sampled explicitly with an additional
level of difficulty in the method \cite{rmp95,como95}. However,
protonic statistics become relevant when the quantum
dispersion is comparable to the interionic distances
$\Lambda_{p}=\sqrt{2\lambda_{p}\beta}\approx
(N_{p}/V)^{-1/3}=n_{p}^{-1/3}$. This define a degeneracy
temperature $k_{B}T_{D}(n_{p})=2\lambda_{p}n_{p}^{2/3}$ below which quantum statistics need to be considered. For
hydrogen $T_{D}\simeq 66.2(K)/r_{s}^{2}$, where
 $r_{s}$ is the usual ion sphere radius of coulomb systems $r_{s}=(3/4\pi n_{p})^{1/3}$.
Therefore proton statistics in metallic hydrogen ($r_{s}\le 1.3$)
becomes relevant below 50K depending on the density, a regime that
we have not investigated yet. Proton statistics in molecular
hydrogen is also quite important and results in the separation
between ortho- and para-hydrogen\cite{Silvera80}. Because this
effect is relevant only at low temperature, we have disregarded
it as well.

In order to implement the PIMC we need a suitable approximation
for the high temperature density matrix $\rho_{p}(S,S'|\tau_{p})$.
We could use either the primitive approximation or the importance
sampling approximation described earlier. However a better
approximation, in particular for distinguishable particles, is the
pair product action \cite{rmp95} which closely resembles the pair
trial function. The idea is to build the many body density matrix
as the product over all distinct pairs of a two-body density
matrices obtained numerically for a pair of isolated particles. At
high temperature the system approaches the classical Boltzmann
distribution which is indeed of the pair product form. The method
is described in detail in ref. \cite{rmp95}. Here we just explain
how we can take advantage of this methodology within the CEIMC
scheme. In order to use the method of pair action, we need to have
a pair potential between quantum particles. In CEIMC, however,
the interaction among protons is provided by the many-body BO
energy. Our strategy is to introduce an effective two-body
potential between protons $\hat{V}_{e}$ and to recast the ionic
density matrix as
\begin{eqnarray}
{\rho}_{_{P}}(S,S'|\tau_{p})&&=<S|e^{-\tau_p[\hat{H}_{e}+(\hat{E}_{BO}-\hat{V}_{e})]}|S'>\nonumber \\
&&\approx
<S|e^{-\tau_p\hat{H}_{e}}|S'> e^{-\frac{\tau_p}{2}[{E}_{BO}(S)-{V}_{e}(S)]+[{E}_{BO}(S')-{V}_{e}(S')]}
\end{eqnarray}
where $\hat{H}_{e}=\hat{K}_{p}+\hat{V}_{e}$ and the corrections
from the effective potential to the true BO energy are treated at
the level of the primitive approximation.
We can compute numerically the matrix elements of the effective pair density matrix
$\hat{\rho}_{e}^{(2)}(\tau_{p})$ as explained in ref. \cite{rmp95}.
The effective $N_{p}$-body density matrix is approximated by \\
\begin{equation}
<S|e^{-\tau_p\hat{H}_{e}}|S'>\approx \prod_{ij}^{N_{p}}<s_{i},s_{j}|\hat{\rho}_{e}^{(2)}
(\tau_{p})|s'_{i},s'_{j}>=\rho_{0}(S,S'|\tau_{p})e^{-\sum_{ij}u_{e}(\vec{s}_{ij},\vec{s}'_{ij}|\tau_{p})}
\end{equation}
where $\rho_{0}$ is the free particle density matrix and
$u_{e}(\vec{s}_{ij},\vec{s}'_{ij}|\tau_{p})$ is the effective pair
action. The explicit form for the partition function is then
\begin{eqnarray}
Z(\beta)=\int dS_{1}\dots dS_{P}&&\prod_{k=1}^{P}
\exp\left\{-\frac{(S_{k}-S_{k+1})^{2}}{4\lambda_{p}\tau_{p}}- \sum_{ij}u_{e}
(\vec{s}^{k}_{ij},\vec{s}^{k+1}_{ij}|\tau_{p})\right\}\nonumber \\
&&\exp\left\{-\tau_{p}\sum_{k=1}^{P}[E_{BO}(S_{k})-V_{e}(S_{k})] \right\}
\end{eqnarray}
with the boundary condition: $S_{P+1}=S_{1}$. As for the
pre-rejection step in the proton moves, we have used different
effective potentials according to the system under consideration.
In metallic hydrogen (a plasma) we have used a smooth screened
coulomb form and found that it provides a fast convergence in
$\tau_{p}$. Convergence can be assessed by monitoring the various
terms in the estimator for the proton kinetic energy\cite{rmp95}.
A good effective potential should provide uniform convergence of
the various orders in $\tau_{p}$. With this effective potential,
we have found convergence to the continuum limit
($\tau_{p}\rightarrow 0$) for $1/\tau_p\ge 3000K$ which allows to
simulate systems at room temperature with only $M\approx10$ proton
slices for $r_s\ge1$. In molecular hydrogen we need to consider
the extra contribution of the bonding potential. We have used the
Kolos-Wolniewicz bonding potential in connection with the same
smooth screened coulomb potential for the non bonding
interactions. Convergence with $\tau_{p}$ is observed in a similar
range.

A very nice feature of ionic PIMC in CEIMC is that considering
ionic paths rather then classical point particles does not add any
computational cost to the method. Let us suppose we run classical
ions with a given level of noise $(\beta\sigma_{cl})^{2}$.
Consider now representing the ions by $P$ time slices. To have a
comparable extra-rejection due to the noise we need a noise level
per slice given by: $(\tau_{p}\sigma_{k})^{2}\approx
(\beta\sigma_{cl})^{2}/P$ which provides $\sigma_{k}^{2}\approx
P\sigma_{cl}^{2}$.
We can allow a noise per time slice P times
larger which means considering P times less independent estimates
of the energy difference per slice. However we need to run P
different calculations, one for each different time slice, so that
the amount of computing for a fixed global noise level is the same
as for classical ions. In practice, however, because our orbitals
depend on all proton positions, we are forced to move the proton
positions at given imaginary time all together with a local (in
imaginary time) update scheme. It is well known that the
autocorrelation time of schemes with local updates rapidly
increases with the chain length and this is the ultimate
bottleneck of our present algorithm \cite{rmp95}. It is therefore essential to
adopt the best factorization in order to minimize the number of
time slices P needed and therefore the efficiency of the method.

When using TABC with quantum ions, for any proton time slice we
should, in principle, perform a separate evaluation of the BO
energy difference averaged over all twist angles. Instead at each
protonic step, we randomly assign a subset of twist phases at each
time slice and we compute the energy difference for that phase
only. The TABC is then performed by adding up all the
contributions from the different time slices. We have checked in
few cases that this simplified procedure does not give
detectable biases in the averages.

In practice we move all slices of all protons at the same time by
a simple random move in a box and we pre-reject the moves with the
effective pair action.

\begin{figure}[htb]
\centering
\includegraphics[scale=0.45]{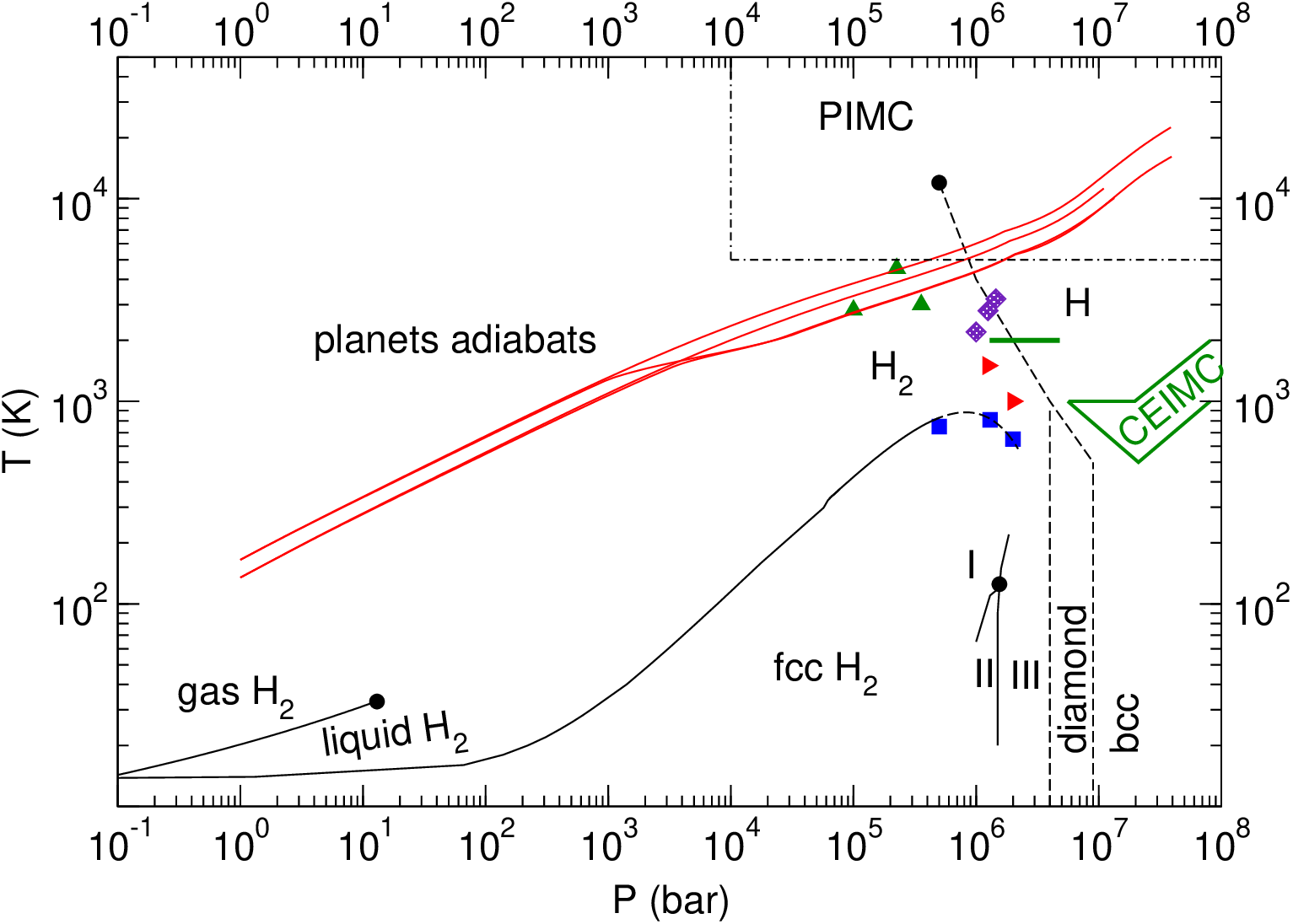}
\caption{Hydrogen phase diagram. Continuous transition lines are experimental results, dashed lines are theoretical prediction from various methods. Squares and right-triangle are ab-initio MD predictions of molecular melting \cite{bonev04} and molecular dissociation in the liquid phase \cite{scandolo03}. The diamonds are shock-waves experimental data through the liquid metalization \cite{WeirMitchellNellis96}. The triangles are earlier CEIMC data for the insulating molecular state \cite{DewingCeperley01, dmc03} while the green domain on the extreme right indicates the CEIMC prediction for the melting \cite{pch04}. Red lines are model adiabats for the interior of the giant planets of the solar system \cite{guillot}. }
\label{fig:PhaseDiagram}
\end{figure}
\section{Summary of results on high pressure hydrogen}
In this section we briefly summarize some of the CEIMC results we
have obtained for high pressure hydrogen. Figure
\ref{fig:PhaseDiagram} shows what is known and what has been
predicted about the hydrogen phase diagram in a wide range of
pressures and temperatures. The rectangular region in the right
upper corner is the region where R-PIMC method can make reliable
predictions \cite{pcbm94,MilitzerCeperley00,MilitzerCeperley01}.
There have been many studies of the ground state of hydrogen (T=0)
including some QMC investigations
\cite{CeperleyAlder87,WangZhuLouieFahy90,natoli93,natoli95}. They have predicted
a metallization density corresponding at $r_{s}=1.31$ accompanied
by a transition from a molecular m-hcp structure to a diamond
lattice of protons. At intermediate temperatures a number of
ab-initio Molecular Dynamics studies have been performed in the
molecular and in the metallic phases, both in the crystal and in
the liquid state \cite{hohl93,kh95,kohanoff97,scandolo03,bonev04}.

\subsubsection{Metallic hydrogen}
We first focus on the metallic system for pressure beyond the
molecular dissociation threshold. In this region, hydrogen is a
system of protons and delocalized electrons. At low enough
temperature the protons order in a crystalline lattice which melts
upon increasing temperature. The low temperature stable structure
as a function of density is still under debate. The most accurate
ground state QMC calculation \cite{natoli93}, indicates
that hydrogen at the edge of molecular dissociation will order in a
diamond structure, and upon increasing density will undergo
various structural transformations ultimately transforming to the
bcc structure. However, these prediction are extrapolated from a
single calculation at $r_{s}=1.31$ and temperature effects are
absent. With CEIMC we have investigated the density range
$r_{s}\in[0.8,1.2]$ and the temperature range $T\in[300K,5000K]$
across the melting transition of the proton lattice and up to the
lower limit of applicability of RPIMC. We limited the study to
systems of $N_{p}=32$ and $N_{p}=54$ which, for cubic simulation
boxes, form fcc and bcc lattices, respectively. We have observed
the melting and refreezing of the protons and made a qualitative
location of the melting line versus the density \cite{dmc03,pch04}
as shown by the green contour lines in figure
\ref{fig:PhaseDiagram}.

\begin{figure}[htb]
\centering
\includegraphics[scale=0.43]{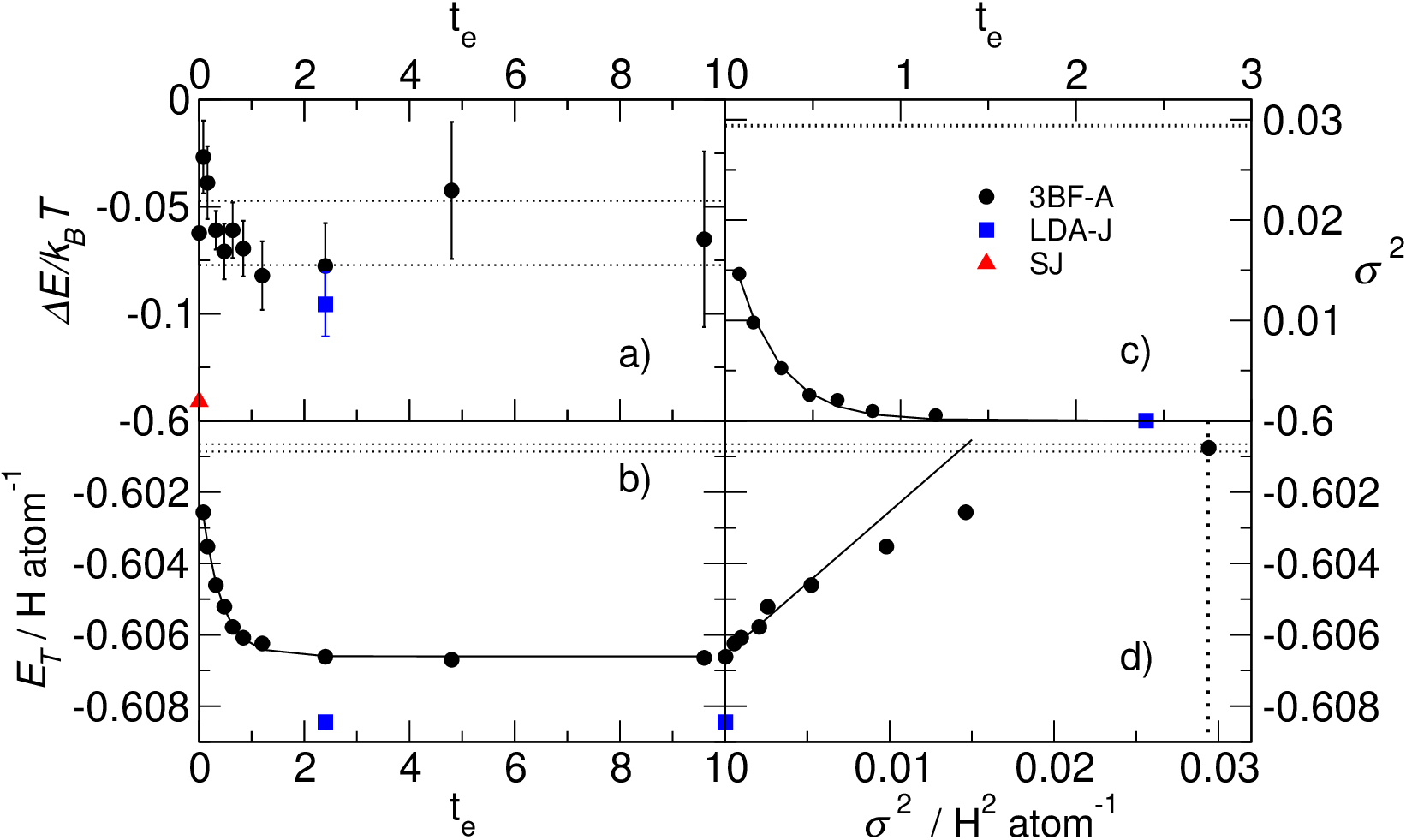}
\caption{$N_{e}=N_{p}=16$, $r_{s}=1.31$. Dependence of total energy, variance
and energy difference for a pair of proton configurations $(S,S')$ on the RQMC projection time. The study is performed for $\tau_{e}=0.02H^{-1}$. Dotted lines
represent the variational estimates with their error bars. In panel
b) and c) the lines are exponential fits to data and in panel d) the
continuous line is a linear fit in the region $\sigma^{2}\leq0.005$. Black circles (3BF-A) are results obtained with the analitical three-body and backflow trial wave functions discussed earlier, the red triangle is a variational result with a Slater-Jastrow trial function with simple plane wave orbitals and the blue squares are results from a trial function with LDA orbitals and an optmized two-body Jastrow.}
\label{fig:figure1}
\end{figure}
A number of interesting question about the convergence and the
efficiency of the CEIMC algorithm need to be answered before
starting a systematic study. An important one is: how large must
the electronic projection time be in order to get convergence in
the energy difference to the ground state value and therefore
obtain unbiased sampling in CEIMC? In order to answer such a
question we have selected a pair of protonic configuration
$(S,S')$ at given density and computed the energy difference,
together with total energy and variance, versus the electronic
projection time. The results reported in the figure
\ref{fig:figure1} correspond to a system of $N_{e}=N_{p}=16$ at
$r_{s}=1.31$ with the twist phase $\vec{\theta}=(0.4, 0.5, 0.6)
\pi$. Results are obtained for an electronic time step
$\tau_{e}=0.02 H^{-1}$, a compromise between accuracy and
efficiency.  In panel a) we report the energy
difference versus the projection time $t$ to show that it does not
depend on the projection time when using the accurate trial
function with 3-body terms and backflow orbitals (3BF-A, black
dots) discussed above. This suggests that the proton configuration
space can be sampled using VMC for the electrons which is faster
and more stable than RQMC. On the same panel we have reported the
VMC estimate obtained with a trial function with a Slater
determinant of simple plane waves and a two-body RPA Jastrow (SJ,
red triangle) and a RQMC result obtained for a trial function with a
2-body RPA-Jastrow and a Slater determinant of self consistent LDA
orbitals (LDA-J, blue squares). The simple SJ function at
the variational level has a much larger energy difference (in
absolute value) and therefore will provide a biased sampling of
protonic configurations. It is not clear whether the RQMC
projection with such trial function will recover the correct
energy difference (remember that the two kind of trial functions
have different nodal structure). On the other hand, no difference
is detected between the RQMC results from the 3BF-A and the LDA-J
trial functions at the same projection time. This gives an
indication of the quality of the 3BF-A nodal surfaces. 
\begin{figure}[]
\centering
\includegraphics[scale=0.45]{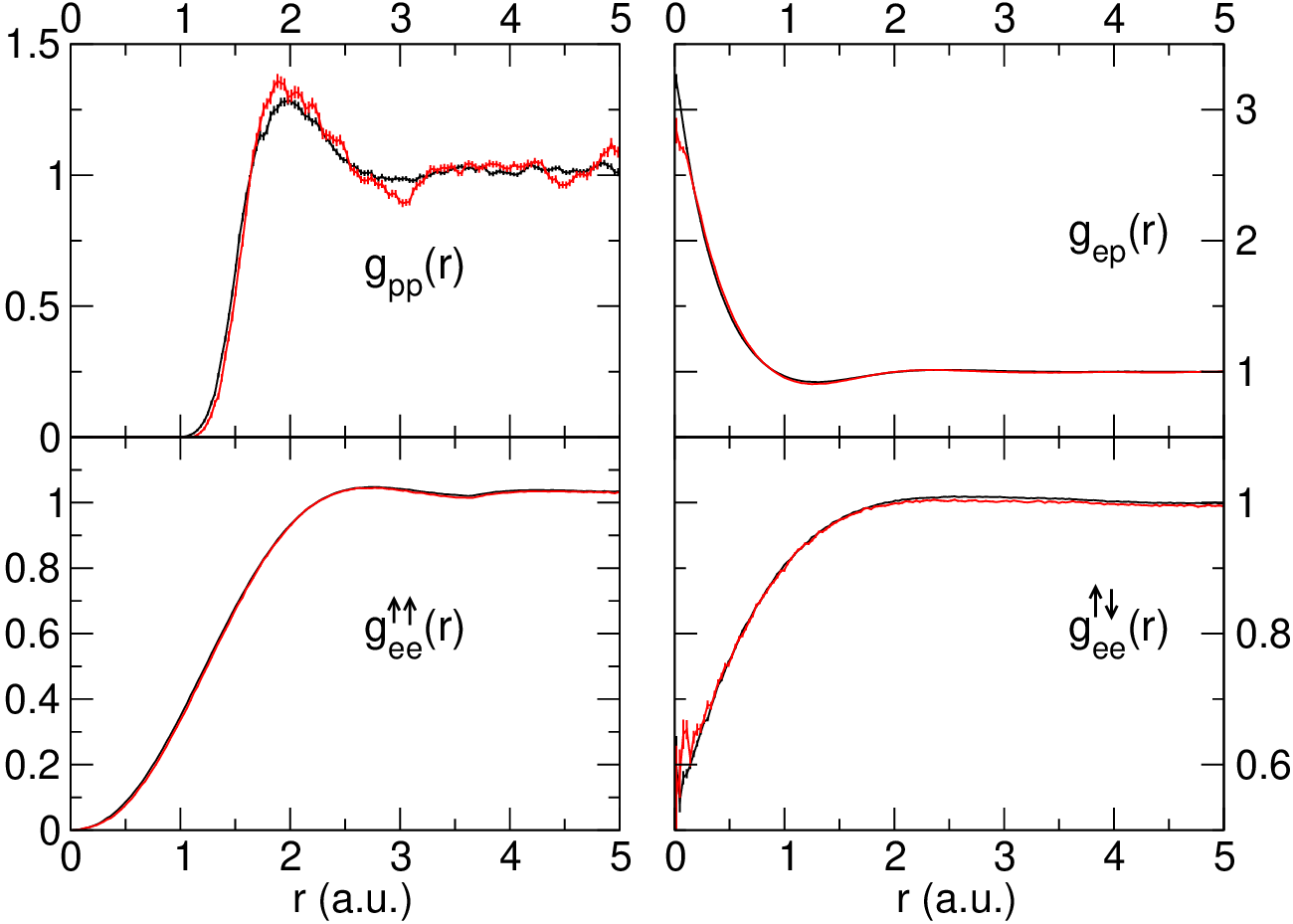}
\caption{$N_{e}=N_{p}=54$, $r_{s}=1.2$, $T=5000K$, $\Gamma$ point. Comparison of VMC (black) and RQMC (red) data for the pair correlation functions. The projection time in RQMC is $t_{e}=0.68H^{-1}$ with an electronic time step of $\tau_{e}=0.01H^{-1}$. Protons are considered as classical point particles. }
\label{fig:figure2}
\end{figure}
The total
energy and the variance of configuration $S$ versus $t$ are
reported in panels b) and c),
respectively. The two panels nicely illustrate the limiting
process toward the fixed-node ground state operated by the
projection (see eq.(\ref{eq:E(t)}) ). Within the 3BF-A trial
function we observe a large gain in energy with the projection
time, the difference being $E(\infty)-E(0)=5.7mH/at=1810K/at$. In
both panels the results of the SJ trial function are off scale,
while we have reported the result of the LDA-J wave function.
According to the total energy, the quality of LDA-J function is
superior to the one of 3BF-A function for this
configuration. The same quality was instead detected for the bcc
lattice configuration \cite{hcpe03}. Finally, 
panel d) we report the total energy versus the variance of the 3BF-A
trial function to show that both quantities go linearly with the
error in the wave function when we are close enough to the ground
state (the fixed-node one) \cite{kwon}. 
\begin{figure}[]
\centering
\includegraphics[scale=0.45]{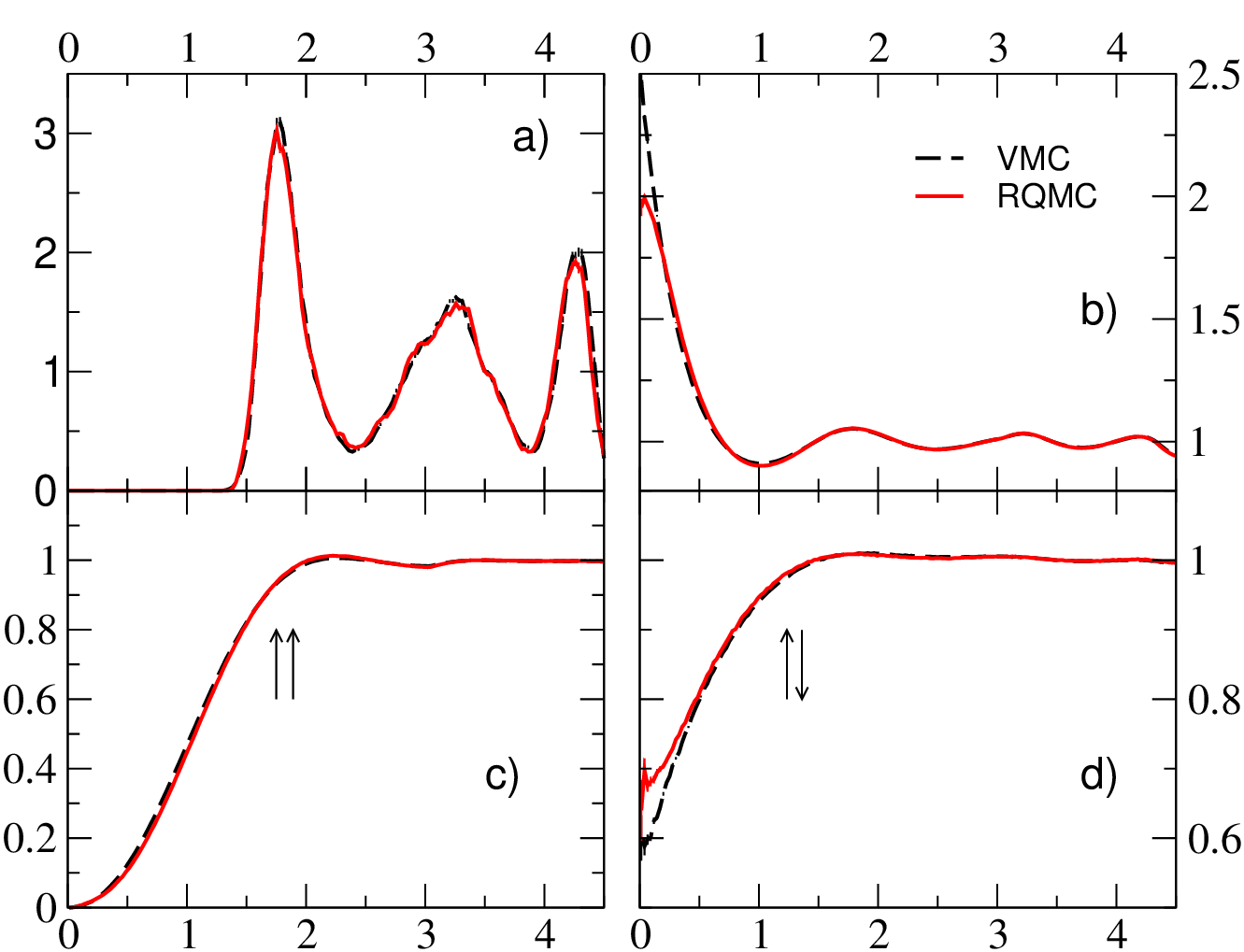}
\caption{$N_{e}=N_{p}=54$, $r_{s}=1.0$, $T=1000K$, $\Gamma$ point. Comparison of VMC (black) and RQMC (red) data for the pair correlation functions. The projection time in RQMC is $t_{e}=1.0 H^{-1}$ with an electronic time step of $\tau_{e}=0.02 H^{-1}$. Protons are considered as classical point particles. The proton-proton pair correlation functions exhibits a structure reminiscent of the bcc crystal structure. RQMC data for $g_{ep}(r)$ and $g_{ee}^{\uparrow\downarrow}(r)$ show a finite time step errors at short distances which however do not contribute significantly to the energy.}
\label{fig:figure3}
\end{figure}
In order to check that we
can indeed sample the proton space with VMC energy differences, we
have performed two test runs for the systems of $N_{p}=N_{e}=54$
at the $\Gamma$ point (zero twist angle), one at $r_{s}=1.2$ and
$T=5000K$ and the other at $r_{s}=1$ and $T=1000K$. The comparison for the pair
correlation functions are shown in figures \ref{fig:figure2} and
\ref{fig:figure3} respectively.
\begin{figure}[]
\centering
\includegraphics[scale=0.45]{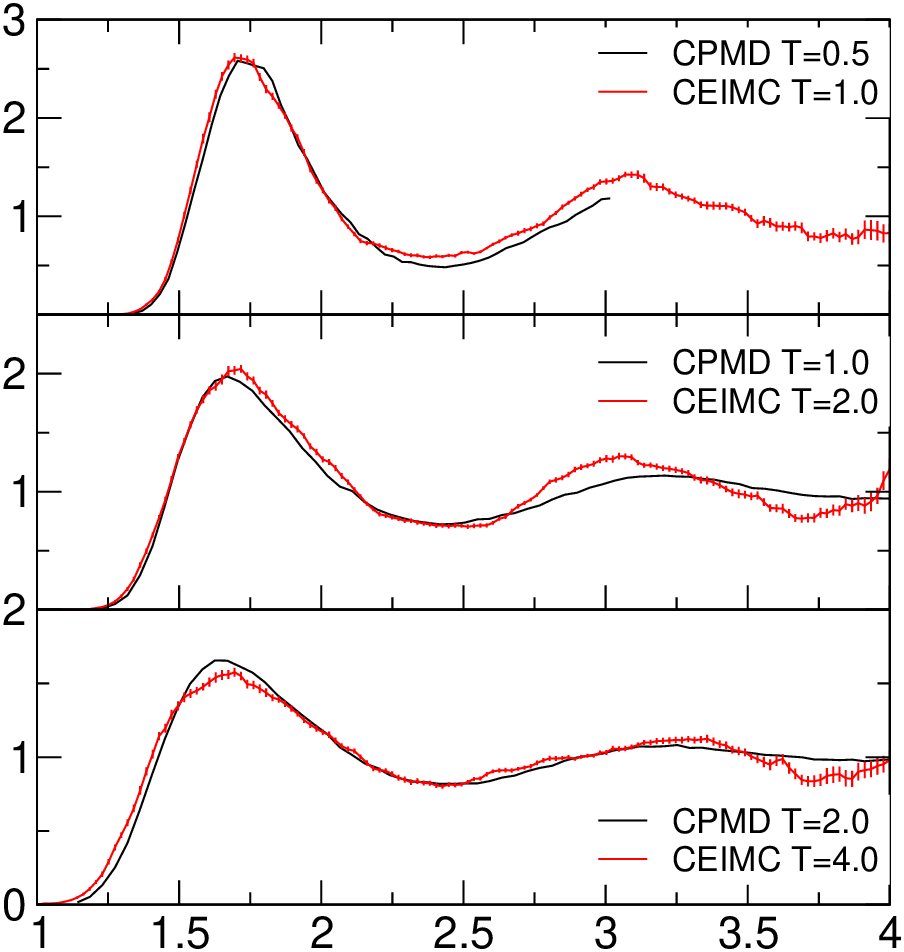}
\caption{Proton-proton pair correlation functions for classical protons at $r_{s}=1$ and various temperatures. Comparison between CEIMC and CPMD-LDA data.}
\label{fig:figure4}
\end{figure}

It is interesting to compare the predictions of CEIMC with other methods. In refs. \cite{dmc03,pch04} we have compared with Restricted Path Integral Monte Carlo data and with Car-Parrinello Molecular Dynamics data with LDA forces (CPMD-LDA) \cite{kh95}. In figure \ref{fig:figure4} we compare CEIMC and CPMD-LDA $g_{pp}(r)$'s for classical protons at $r_{s}=1$ and various temperatures. Both calculations are done with PBC (zero twist angle) and CEIMC uses 54 protons while CPMD-LDA used 162 protons (both are closed shells in the reciprocal space, so that the electronic ground state is not degenerate). We see that in a wide range of temperatures, CPMD-LDA and CEIMC are off by a factor of 2 in temperature. CPMD-LDA predicts a less structured fluid and locates the melting transition at roughly $T_{m}\simeq 350K$ \cite{kh95}. With CEIMC instead more structure is found and the melting is located between 1000K and 1500K \cite{pch04}. Though several reasons could be at the origin of such unexpected discrepancy, we believe that the probelm arises from a too smooth BO energy surface as provided by LDA. Evidences of this fact are also provided by previous ground state calculations for hydrogen crystal structures \cite{natoli93} which similarly found energy differences between various structures from LDA to be roughly half of the corresponding one from DMC. This will explaine the factor of 2 in the temperature scale observed in figure \ref{fig:figure4}.

In ref. \cite{pch04} we have reported data for the equation of
state of metallic hydrogen. Proton quantum effects are quite
important at such high density and need to be considered carefully
to get accurate prediction of the equation of state. The
importance of proton quantum effect is partially reported in ref.
\cite{hpc05}. Also we have shown that electronic VMC with 3BF-A
trial function is accurate enough to sample the proton
configuration space. However RQMC should be used in order to
obtain accurate results for the energy and the pressure. A good
strategy is to run CEIMC with VMC to sample efficiently the proton
configurations and then run RQMC on fixed statistical
independent protonic configurations previously generated.

\subsubsection{Insulating Molecular Hydrogen}
In this section we report very preliminary results in the insulating molecular liquid phase. 
We have investigated the state point at $r_{s}=2.1$ and $T=4530K$
because at this point, and other scattered points around it,
experimental data for the equation of state are available from
Gas-gun techiniques \cite{holmes95}.\begin{figure}[]
\centering
\includegraphics[scale=0.45]{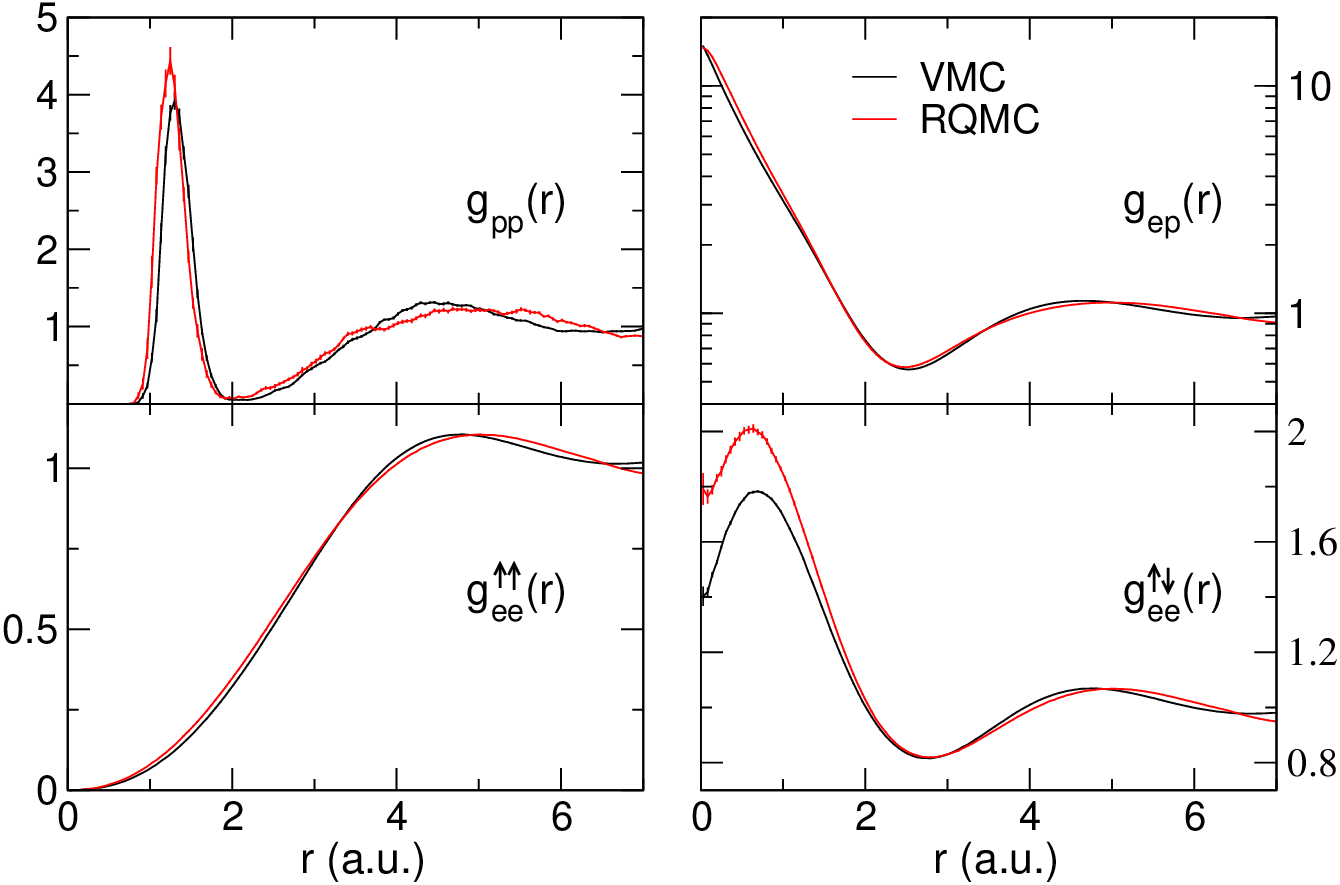}
\caption{Comparison between VMC and RQMC computed pair correlation functions for insulating molecular hydrogen at $r_{s}=2.1, T=4530K$. $N_{p}=N_{e}=54$.}
\label{fig:Molecular}
\end{figure}
 Also the first CEIMC
attempt \cite{DewingCeperley01,dmc03} focus on the same point although
using quite a different trial functions. We used a pair product
trial function with modified RPA Jastrows (electron-electron and
electron-proton) and Slater determinants built with the molecular
orbitals of eq. (\ref{eq:MolecularOrbitals}). At variance with the
metallic trial function, we have now 5 variational parameters to
be optimized, 4 in the Jastrow factor and a single one in the orbitals.
In particular, the nodal structure of the wave function will be
affected by the width of the gaussian orbitals used to build the
molecular orbitals. In the first CEIMC attempt \cite{DewingCeperley01,dmc03}, a single
multivariate gaussian, centered at the molecular center of mass
and with a width different for each molecule was used. Thus the
number of variational parameter was equal to three time the number
of molecules. Also, the entire optimization procedure for each
attempted protonic configuration was performed prior the
Metropolis test. Clearly, the optimization step was a bottleneck
of that scheme. In our present scheme, we have gained evidence
that we can optimize the parameters on a single protonic
configuration (even  the parameters optimized on a lattice configuration are
good) and use the same values for simulating the liquid phase. A
similar conclusion was obtained in the metallic phase with
numerically optimized orbitals \cite{dmc03}. In this way the
optimization step needs to be performed only upon changing density
and/or number of particles, while we can use the same set of
values for the variational parameters to span the temperature axis
at fixed density and number of particles. In figure
\ref{fig:Molecular} we compare the pair correlation functions as
obtained by VMC and RQMC with $\tau_{e}=0.01 H^{-1}$ and
$t_{e}=1.6 H^{-1}$ for a system of 27 hydrogen molecules. Since
the system is insulating we do not average over the twist angle,
but just use the $\Gamma$ point. We observe that RQMC enhances
slightly the stregth of the molecular bond and the electronic
correlation inside molecules with respect to VMC. Good agreement
with the early CEIMC results is observed not
only for the correlation function, but also for the equation of
state. In our present calculation we obtain
$P(RQMC)=0.224(5)Mbars$ to be compared with the previous estimate
of $0.225(3)Mbars$  \cite{DewingCeperley01,dmc03} and with the experimental data $P=0.234Mbars$.
The deviation from the experimental data is only roughly $5\%$ and
it is particularly encouraging if we consider that our data are
for classical protons and quantum effects are expected to slightly
increase the pressure.

\section{Conclusions and future developments}
In this paper we have described the principles of CEIMC and given some technical details on its practical implementation. The results for metallic and molecular hydrogen show that CEIMC is a practical strategy to couple ground state QMC methods for the electronic degrees of freedom with a finite temperature Monte Carlo simulation of the ionic degrees of freedom. To our knowledge, CEIMC is the only method available so far to perform ab-initio simulations based on QMC energies. 

In a recent work, Grossman and Mitas have proposed a strategy to correct ab-initio Molecular Dynamics energies with QMC \cite{GrossmanMitas05}. This is however different from CEIMC because the nuclear degrees of freedom are still sampled on the basis of DFT forces. Therefore, when applied to metallic hydrogen, that method would have found the same liquid structure as obtained by CPMD-LDA \cite{kh95}. 

Very recently an interesting proposition on how to use noisy forces in ab-initio Molecular Dynamics has appeared \cite{KrajewskiParrinello05}. The same strategy could be used with noisy QMC forces to simulate the dynamics of classical ions. A first attempt has already shown the feasibilty of this promising method \cite{attaccalite} although the results are still very preliminary. 

A crucial aspect of CEIMC is the choice of the electronic trial wave function. The ones we have discussed in this paper are either suitable for the metallic state or for the molecular state and their quality in describing the metalization-molecular dissociation transition is questionable. A current development within CEIMC is an efficient strategy to generate on the fly single electron orbitals depending on the instantaneous ionic configuration, in the spirit of the LDA orbitals previously used \cite{natoli93}. We have devised an efficient algorithm which is able to provide accurate orbitals in a reasonable computer time. We are at present using these orbitals to explore the molecular dissociation under pressure in the liquid state \cite{DelaneyCeperleyPierleoni_tbp}. The same strategy can be used to obtain accurate prediction for the hydrogen equation of state in the temperature range between 50K and 5000K. Also application to helium and helium-hydrogen mixtures, very interesting systems for planetary physics, can be envisaged with the same methodology.

Tests for non-hydrogenic systems are needed to find the
performance of the method on a broader spectrum of
applications. The use of pseudopotentials within QMC to treat
atoms with inner core is well tested \cite{Foulkes01}. What is not clear is how
much time will be needed to generate trial functions, and to
reduce the noise level to acceptable limits. Clearly, further work is needed to allow this next step in the development of microscopic simulation algorithms.

\printindex
\end{document}